\documentclass[11pt,a4paper,reqno]{amsart}
\usepackage{a4wide}
\usepackage{amsfonts,amsmath,amssymb,amsthm,enumerate,color,hyperref,tabularx,ifthen,twoopt,booktabs,multirow}
\usepackage{url,natbib,graphicx}

\newcommand{\PE}{\mathbb{E}}

\begin{document}

\title[A multivariate variable selection approach for LC-MS metabolomics data]{A multivariate variable selection approach for analyzing LC-MS metabolomics data}

\author[M. Perrot-Dock\`es, C. L\'evy-Leduc, J. Chiquet et al.]{M. Perrot-Dock\`es, C. L\'evy-Leduc, J. Chiquet, L.  Sansonnet, 
M. Br\'eg\`ere, M.-P. \'Etienne, S. Robin}
\address{UMR MIA-Paris, AgroParisTech, INRA, Universit\'e 
Paris-Saclay, 75005, Paris, France}
\email{marie.perrot-dockes@agroparistech.fr}
\author[]{G. Genta-Jouve}
\address{UMR CNRS 8638 Com\`ete - Universit\'e Paris-Descartes, CNRS, 75006 Paris France.}

\begin{abstract} Omic data are characterized by the
    presence of  strong dependence structures that  result either from
    data  acquisition or  from  some underlying  biological
    processes.   In metabolomics,  for  instance,  data resulting  from
    Liquid  Chromatography-Mass Spectrometry  (LC-MS)  -- a  technique
    which gives access  to a large coverage of  metabolites -- exhibit
    such  patterns.   These  data sets  are   typically  used  to  find  the
    metabolites characterizing a phenotype of interest associated with
    the samples. However, applying
    some  statistical  procedures that  do  not  adjust the variable selection
    step to  the dependence pattern may result  in a loss
    of power  and the selection  of spurious variables.  The  goal of
  this  paper is  to propose  a  variable selection  procedure in  the
  multivariate linear model that accounts for the dependence structure
  of  the multiple  outputs which may lead in the LC-MS framework to the selection
 of more relevant metabolites.
    We propose  a novel  Lasso-based approach  in the
  multivariate  framework  of the  general  linear  model taking  into
  account the dependence structure by  using various modelings of the
  covariance matrix  of the residuals. Our  numerical experiments show
  that  including  the estimation  of  the  covariance matrix  of  the
  residuals in the Lasso  criterion dramatically improves the variable
  selection performance. Our approach  is also successfully applied to
  a LC-MS data set made of African  copals samples for which it is able
  to provide a small list of metabolites without altering the
  phenotype discrimination.
   Our methodology is implemented in the R package \textsf{MultiVarSel}
which is available from the CRAN (Comprehensive R Archive Network).
\end{abstract}

\maketitle


 \section{Introduction}

Metabolomics aims to provide a global snapshot (quantitative or
qualitative) of the metabolism at a given time and by extension, a
phenotypic information, see \cite{Nicholson1999}. 
To this end, it studies the concentration in small molecules called metabolites
that are the end products of the enzymatic machinery of the cell.
Indeed, minor variations in gene or protein expression levels that are
not observable via high throughput experiments may have an influence
on the metabolites and hence on the phenotype of interest. Thus,
metabolomics is a promising approach that can advantageously
complement usual transcriptomic and proteomic analyses.

In metabolomics, the analysis  of the biological samples is often performed
using High Resolution Mass Spectrometry (HRMS), Nuclear Magnetic Resonance (NMR)  or
Liquid  Chromatography-Mass Spectrometry  (LC-MS)   
and produces a large number of features (hundreds or thousands) 
that can explain a difference between two or more
populations, see \cite{Zhang2012}. It is well-known that the identification
of metabolites discriminating these populations remains a major bottleneck in metabolomics and therefore 
the selection of relevant features (variables) is a crucial step in the
statistical analysis of the metabolomic data, as explained in \cite{Verdegem2016}.

Different supervised machine learning approaches have been  used in
metabolomics during the last few years, see \cite{Saccenti2013,
  Ren2015, Boccard201618}. Among them the most widely used is 
the partial least squares-discriminant analysis (PLS-DA) which has recently been extended to sPLS-DA (sparse partial least squares-discriminant analysis)
by \cite{LeCao2011} to include a variable selection step. 
Nevertheless, \cite{Grissa2016}
 highlight the need for new development in the process of
 features selection that would take into account the specificity of
 metabolomics data which is the dependence that may exist between the
 different metabolites. In this perspective, our paper proposes a novel feature selection
 methodology which consists in a
 variable selection approach based on the Lasso criterion in a
 multivariate setting taking into account the dependence that may exist between the
 different metabolites.

More precisely, let us consider a classical metabolomics experiment
where $n$ samples have been collected and analyzed. This results in an $n \times q$ data matrix where $q$ stands for the number of metabolites.
When the $n$ samples have been obtained under various conditions, we
are typically interested in understanding the effect of each condition
on each metabolite. In the case where $C$ experimental conditions are
compared, $n_c$ denotes the number of replicates under condition $c$, where
$c \in \{1, \dots C\}$ and $\sum_{c=1}^Cn_c = n$. 
We further denote $Y_{c, r}^{(j)}$ the centered LC-MS signal obtained
for the $j$th metabolite ($j \in \{1, \dots q\}$) under Condition $c$ 
for Replicate $r$ ($r \in \{1, \dots n_c\}$). 
In the following, the set of conditions will be called the
``factor'', each specific condition being a ``level'' of this factor. 
The most popular model to analyze quantitative observations $Y$ as a
function of a qualitative variable, that is a factor,
is the analysis of variance (ANOVA) model, which we write here as follows:
\begin{equation}\label{eq:model:anova:simple}
 Y_{c, r}^{(j)} = \mu_c^{(j)} + E_{c, r}^{(j)},
\end{equation}
where the observations $\{Y_{c, r}^{(j)}\}$ are assumed to be centered, so that $\mu_c^{(j)}$ can be interpreted as the effect of Condition $c$ (Level $c$) on
Metabolite $j$ and where the residual terms $\{E_{c, r}^{(j)}\}$ are
assumed to be independent and identically distributed (i.i.d.)
zero-mean Gaussian random variables. 
The goal of such a modeling is to highlight which effects among the
$\mu_1^{(j)},\mu_2^{(j)},\dots,\mu_C^{(j)}$ are the most significant
for the metabolite $j$ since the $\{Y_{c, r}^{(j)}\}$ are assumed to be centered.

When the whole $n \times q$ data matrix is considered instead of a
  single column $j$, 
the model can be summarized in the following matrix form:
\begin{equation}\label{eq:model:matriciel}
\boldsymbol{Y}=\boldsymbol{X}\boldsymbol{B}+\boldsymbol{E},
\end{equation}
where $\boldsymbol{Y}=(Y_{i,j})_{1\leq i\leq n,\; 1\leq j\leq q}$ is
the $n\times q$ observation matrix, $\boldsymbol{X}$ is the $n\times p$ design matrix,
$\boldsymbol{B}$ is the $p\times q$ coefficient matrix and
 $\boldsymbol{E}=(E_{i,j})_{1\leq i\leq n,\; 1\leq j\leq q}$ is the
 $n\times q$ matrix of residual errors. Observe that $p$
 corresponds to the number of explicative variables,
   which is simply $C$ in Model~\eqref{eq:model:anova:simple}.
For notational simplicity, the indices $c,r$ in $Y_{c,r}^{(j)}$ are
summarized in a unique index $i$ in $\{1, \dots n\}$.

In this paper, we pay a special attention to the potential dependence that may exist
among the columns of $\boldsymbol{Y}$, namely the different metabolites. To this aim, we shall assume that
for each $i$ in $\{1,\dots,n\}$,
\begin{equation}\label{eq:def_E}
(E_{i,1},\dots,E_{i,q})\sim\mathcal{N}(0,\boldsymbol{\Sigma}_q),
\end{equation}
where $\boldsymbol{\Sigma}_q$ denotes the covariance matrix of the
$i$th row of the residual error matrix. Note that the model defined
by (\ref{eq:model:matriciel}) and (\ref{eq:def_E}) is usually called 
a general linear model or a multivariate linear model which 
has been extensively studied in \cite{mardia:kent:1979}.

The simplest assumption regarding the dependence structure of the
noise is $\boldsymbol{\Sigma}_q=\sigma^2\boldsymbol{I}_q$, where $\boldsymbol{I}_q$
denotes the $q\times q$ identity matrix. In this case the different columns of $\boldsymbol{Y}$ are assumed to be independent.
In more general cases, the matrix
$\boldsymbol{\Sigma}_q$ models the dependence between the different
columns of $\boldsymbol{Y}$, namely the dependence between the metabolites.
In the following, we shall moreover assume that
$(E_{i,1},\ldots,E_{i,q})$ and $(E_{k,1},\ldots,E_{k,q})$ are independent, when $i\neq k$, which
means that the individuals are assumed to be independent.

The problem of finding which parameters are significant among the $(\mu_c^{(j)})_{1\leq
  c\leq C,1\leq j\leq q}$  in Model~(\ref{eq:model:anova:simple})
boils down to finding the non null coefficients in the matrix
$\boldsymbol{B}$ in Model (\ref{eq:model:matriciel}) and hence can be seen as a variable selection problem
in the general linear model.
Several approaches can be considered for solving this issue: either 
\textit{a posteriori} methods such as classical statistical tests in ANOVA
models, see \cite{mardia:kent:1979,faraway:2005}
or methods embedding the variable selection such as Lasso-type
methodologies initially proposed by \cite{Tib96}. However, a raw application of such approaches does not 
take into account the potential dependence between the
different columns of $\boldsymbol{Y}$. This drawback will be illustrated in Section
\ref{sec:num_exp}.

The goal of our paper is twofold: First, to remedy the limitations of these approaches
by proposing a method for estimating the dependence between the columns
of $\boldsymbol{Y}$ and second, to deal with the potentially high number of
variables by using a Lasso-type approach taking into
  account this dependence. For this purpose, we shall propose a three-step inference
strategy further detailed hereafter.

The paper is organized as follows. Our method is described in Section \ref{sec:stat_inf}.
To support our methodology, some numerical experiments on synthetic data are provided in Section \ref{sec:num_exp}.
Finally, an application to a metabolomics data set produced by the analysis of African
copals samples is given in Section \ref{sec:real}.


\section{Statistical inference}\label{sec:stat_inf}

 The strategy that we propose can be summarized as follows.
\begin{itemize}
\item \textsf{First step:} Fitting a one-way ANOVA to each column of the observation matrix 
  $\boldsymbol{Y}$ in order to have access to an
  estimation $\widehat{\boldsymbol{E}}$ of the residual matrix
  $\boldsymbol{E}$.
\item \textsf{Second step:} Estimating the matrix
  $\boldsymbol{\Sigma}_q$ by using the methods described in Sections
  \ref{subsec:param} and \ref{subsec:nonparam}. Then, choosing the
  most convenient estimator $\widehat{\boldsymbol{\Sigma}}_q$
  thanks to a statistical test described in Section \ref{sec:whitening_test}.
\item \textsf{Third step:} Thanks to the estimator $\widehat{\boldsymbol{\Sigma}}_q$, transforming the
  data in order to remove the dependence between the columns of
  $\boldsymbol{Y}$. Such a transformation will be called ``whitening'' hereafter.
  Then, applying to these
  transformed observations the Lasso approach described in Section \ref{sec:lasso}.
\end{itemize}

The first step provides  a first estimate $\widetilde{\boldsymbol{B}}$
of $\boldsymbol{B}$. An estimate of the residual  matrix is then
defined                                                             as
$\widehat{\boldsymbol{E}}    =    \boldsymbol{Y}   -    \boldsymbol{X}
\widetilde{\boldsymbol{B}}$.
In the following, we shall focus on the last two steps.

\subsection{Estimation of the dependence structure of $\boldsymbol{E}$}\label{sec:estim_sigma_q}

We propose hereafter to model each row of $\boldsymbol{E}$ as realizations 
of a stationary process and hence we shall use time-series models in order
to  describe the  dependence structure of  $\boldsymbol{E}$. We refer the reader to \cite{brockwell:davis} for further
details on time series modeling.   

We  shall consider  hereafter  a  large variety  of  models
ranging from the simplest parametric to the most general nonparametric
dependence  modeling.  In  each case  we  focus on  the estimation  of
$\boldsymbol{\Sigma}_q^{-1/2}$     since    using     the    following
transformation:
\begin{equation}\label{eq:modele:blanchi_est}
\boldsymbol{Y}\;\boldsymbol{\Sigma}_q^{-1/2}=\boldsymbol{X}\boldsymbol{B}\;\boldsymbol{\Sigma}_q^{-1/2}
+\boldsymbol{E}\;\boldsymbol{\Sigma}_q^{-1/2}
\end{equation}
removes the dependence between the columns of $\boldsymbol{Y}$. Such a procedure will be called ``whitening'' hereafter.

\subsubsection{Parametric dependence}\label{subsec:param}
The simplest model among the parametric models is the autoregressive process
of order 1 denoted AR(1). More precisely, for each $i$ in
$\{1,\dots,n\}$, $E_{i,t}$ satisfies the following equation:
\begin{equation}\label{eq:AR1}
E_{i,t}-\phi_1 E_{i,t-1}=W_{i,t},\textrm{ with } W_{i,t}\sim WN(0,\sigma^2),
\end{equation}
where $\phi_1$ is a real number and $WN(0,\sigma^2)$ denotes a zero-mean white noise process of variance $\sigma^2$, namely, if $Z_{t}~\sim~WN(0,\sigma^2)$, then
$\PE(Z_t)=0$, $\PE(Z_t Z_{t'})=0$ if $t\neq t'$ and $\PE(Z_t^2)=\sigma^2$.

In this case, the inverse of the square root of the covariance matrix
$\boldsymbol{\Sigma}_q$ of $(E_{i,1},\dots,E_{i,q})$ has a simple closed-form 
expression given by
\begin{equation}\label{eq:Sigmaq-1/2}
\boldsymbol{\Sigma}_q^{-1/2} =\left(
\begin{matrix}
\sqrt{1-\phi_1^2} & -\phi_1 & 0 & \cdots & 0\\
0 & 1 & -\phi_1 & \cdots & 0 \\
0 & 0 & \ddots & \ddots & \vdots \\
\vdots & \vdots &  \ddots & \ddots & -\phi_1 \\
0 & 0 & \cdots  & 0 & 1 &
\end{matrix}
\right).
\end{equation}
Hence, to obtain the expression of $\widehat{\boldsymbol{\Sigma}}_q^{-1/2}$, it is enough to have an estimation of the parameter $\phi_1$
and to replace it in (\ref{eq:Sigmaq-1/2}). For this, we use the
estimator $\widehat{\boldsymbol{E}}$ of the residual errors matrix
obtained by fitting a standard ANOVA model to the observations, which
corresponds to the first step of our method. Then
$\phi_1$ is estimated by $\widehat{\phi}_1$  defined by
$$
\widehat{\phi}_1=\frac1n\sum_{i=1}^n \widehat{\phi}_{1,i},
$$
where $\widehat{\phi}_{1,i}$ denotes the estimator of $\phi_1$ obtained by the classical Yule-Walker equations from 
$(\widehat{E}_{i,1},\dots,\widehat{E}_{i,q})$, see \cite{brockwell:davis} for more
details. 

More generally, it is also possible to have access to
$\boldsymbol{\Sigma}_q^{-1/2}$ for more complex processes such as the ARMA($p,q$) process defined
as follows: For each $i$ in $\{1,\dots,n\}$,
\begin{equation}\label{eq:ARMA}
E_{i,t}-\phi_1 E_{i,t-1}-\dots-\phi_p E_{i,t-p} =W_{i,t}+\theta_1
W_{i,t-1}+\dots\theta_q W_{i,t-q},
\end{equation}
where $W_{i,t}\sim WN(0,\sigma^2)$, the $\phi_i$'s and the $\theta_i$'s are real numbers.

\subsubsection{Nonparametric dependence case}\label{subsec:nonparam}

In the situation where a parametric modeling is not relevant for
$\boldsymbol{\Sigma}_q$, it can be estimated by
\begin{equation}
\widehat{\boldsymbol{\Sigma}}_q=\left(
\begin{matrix}
\widehat{\gamma}(0) & \widehat{\gamma}(1) & \cdots & \widehat{\gamma}(q-1)\\
\widehat{\gamma}(1) & \widehat{\gamma}(0) & \cdots & \widehat{\gamma}(q-2)\\
\vdots & & & \\
\widehat{\gamma}(q-1) & \widehat{\gamma}(q-2) & \cdots & \widehat{\gamma}(0)
\end{matrix}
\right),
\end{equation}
with
$$
\widehat{\gamma}(h)=\frac1n\sum_{i=1}^n \widehat{\gamma}_{i}(h),
$$
where $\widehat{\gamma}_{i}(h)$ is the standard autocovariance
estimator of $\gamma_i(h)=\PE(E_{i,t}E_{i,t+h})$, for all $t$. Usually, $\widehat{\gamma}_{i}(h)$ is 
referred to as the empirical autocovariance of the
$\widehat{E}_{i,t}$'s at lag $h$  (\textit{i.e.}\,the
  empirical covariance between 
$(\widehat{E}_{i,1}, \dots ,\widehat{E}_{i,n-h})$ and $(\widehat{E}_{i,h+1}, \dots ,\widehat{E}_{i,n})$).
For a definition of the standard autocovariance estimator we
refer the reader to Chapter 7 of \cite{brockwell:davis}.
The matrix $\widehat{\boldsymbol{\Sigma}}_q^{-1/2}$ is then obtained by inverting the Cholesky factor of $\widehat{\boldsymbol{\Sigma}}_q$.

\subsubsection{Choice of the whitening modeling}\label{sec:whitening_test}

In order to decide which dependence modeling is the most adapted to
the data at hand we propose hereafter a statistical test.
If the whitening modeling used is well chosen then each row of 
$\widetilde{\boldsymbol{E}}=\widehat{\boldsymbol{E}}\widehat{\boldsymbol{\Sigma}}_q^{-1/2}$
should be a white noise. 

One of the most popular approach for testing whether a random process
is a white noise is the Portmanteau test which is based on the
Bartlett theorem, for further details we refer the reader to
\cite[Theorem 7.2.2]{brockwell:davis}. By this theorem, we get that
under the null hypothesis $(H_0)$: ``For each $i$ in $\{1,\dots,n\}$,
$(\widetilde{E}_{i,1},\ldots,\widetilde{E}_{i,q})$ is a white noise'',
\begin{equation}\label{eq:stat_test_1}
q\sum_{h=1}^H \widehat{\rho}_i(h)^2\approx \chi^2(H),\textrm{ as }
q\to\infty,
\end{equation}
for each $i$ in $\{1,\dots,n\}$,
where $\widehat{\rho}_i(h)$ denotes the empirical autocorrelation of
$(\widetilde{E}_{i,t})_t$ at lag $h$ and $\chi^2(H)$ denotes the
chi-squared distribution with $H$ degrees of freedom. Thus, by (\ref{eq:stat_test_1}),
we have at our disposal a $p$-value for each $i$ in $\{1,\dots,n\}$
that we denote by $\textrm{Pval}_i$.
In order to have a single $p$-value instead of $n$, we shall consider 
\begin{equation}\label{eq:stat_test_2}
q\sum_{i=1}^n\sum_{h=1}^H \widehat{\rho}_i(h)^2\approx \chi^2(nH),\textrm{ as }
q\to\infty,
\end{equation}
where the approximation comes from the fact that the rows of $\widetilde{\boldsymbol{E}}$
are assumed to be independent. Equation (\ref{eq:stat_test_2}) thus provides a $p$-value: $\textrm{Pval}$.
Hence, if $\textrm{Pval}<\alpha$, the null hypothesis $(H_0)$ is
rejected at the level $\alpha$, where $\alpha$ is usually equal to 5\%. In such a situation taking the dependence into account and estimating the dependence
by one of the previous methods should highly improve the modeling and the variable selection step.

\subsection{Estimation of $\boldsymbol{B}$}\label{sec:lasso}

\subsubsection{Lasso based approach}

Let us first explain briefly the usual framework in which the Lasso
approach is used. We consider a high-dimensional linear model of the following form
\begin{equation}\label{eq:model_vec}
\mathcal{Y}=\mathcal{X}\mathcal{B}+\mathcal{E},
\end{equation}
where $\mathcal{Y}$, $\mathcal{B}$ and $\mathcal{E}$ are vectors. Note that, in
high-dimensional linear models, the matrix $\mathcal{X}$ has usually more columns than rows which means that
the number of variables is larger than the number of observations but
$\mathcal{B}$ is usually a sparse vector, namely it contains a lot of
null components.

In such models a very popular approach initially proposed by \cite{Tib96} consists in using the Least Absolute
Shrinkage eStimatOr (LASSO) criterion for estimating $\mathcal{B}$ defined as follows for a positive $\lambda$:
\begin{equation}\label{eq:lasso}
\widehat{\mathcal{B}}(\lambda)=\textrm{Argmin}_\mathcal{B}\left\{\|\mathcal{Y}-\mathcal{X}\mathcal{B}\|_2^2+\lambda\|\mathcal{B}\|_1\right\},
\end{equation}
where, for $u=(u_1,\dots,u_n)$, $\|u\|_2^2=\sum_{i=1}^n u_i^2$ and $\|u\|_1=\sum_{i=1}^n |u_i|$, which is usually called the $\ell_1$-norm of the vector $u$. Observe that the first term of (\ref{eq:lasso})
is the classical least-squares criterion and that $\lambda\|\mathcal{B}\|_1$ can be seen as a penalty term. The interest of such
a criterion is the sparsity enforcing property of the  $\ell_1$-norm ensuring that the number of non-zero components of the estimator 
$\widehat{\mathcal{B}}$ of $\mathcal{B}$ is small for large enough values
of $\lambda$.
Such a criterion is very relevant in our framework since the problem of finding the significant variables
boils down to finding the non null coefficients in the matrix
$\boldsymbol{B}$.

This methodology cannot be directly applied to our model since we have
to deal with  matrices and not with vectors. However,  as explained in
Appendix A,  Model (\ref{eq:model:matriciel})  can be rewritten  as in
\eqref{eq:model_vec}    where    $\mathcal{Y}$,   $\mathcal{B}$    and
$\mathcal{E}$   are  vectors   of  size   $nq$,  $pq$   and  $nq$,
respectively.  Hence,  retrieving  the   positions  of  the  non  null
components in $\mathcal{B}$  is a first approach  for finding relevant
variables.

However, this approach does not take into account the dependence between the columns of 
$\boldsymbol{Y}$. Hence,
we propose hereafter a modified version of the standard Lasso criterion (\ref{eq:lasso}) taking into account this potential dependence.

As explained previously, our contribution consists first in ``whitening'' the observations, namely removing the dependence that may exist within the observations
matrix, by multiplying (\ref{eq:model:matriciel}) on the right by
$\widehat{\boldsymbol{\Sigma}}_q^{-1/2}$, see
(\ref{eq:modele:blanchi_est}) where $\boldsymbol{\Sigma}_q^{-1/2}$ is
replaced by $\widehat{\boldsymbol{\Sigma}}_q^{-1/2}$. 
By using the same vectorization trick that allows us to transform Model (\ref{eq:model:matriciel}) into Model (\ref{eq:model_vec}),
the Lasso criterion can be applied to the vectorized version of Model
(\ref{eq:modele:blanchi_est}) where $\boldsymbol{\Sigma}_q^{-1/2}$ is
replaced by $\widehat{\boldsymbol{\Sigma}}_q^{-1/2}$.
The specific expressions of $\mathcal{Y}$, $\mathcal{X}$, $\mathcal{B}$ and $\mathcal{E}$ are given in Appendix B.

Note that this idea of ``whitening'' the observations
has also been proposed by \cite{rothman:2010} in which the estimation of $\boldsymbol{\Sigma}_q$ and $\boldsymbol{B}$
is performed simultaneously. An implementation is available in the R package \textsf{MRCE}.
In our approach, $\boldsymbol{\Sigma_q}$ is estimated first and then
its estimator is used in (\ref{eq:modele:blanchi_est}) instead of 
$\boldsymbol{\Sigma}_q$ before applying the Lasso criterion. Hence, our method can be seen as a variant of the MRCE method
in which $\boldsymbol{\Sigma}_q$ is estimated beforehand. Moreover,
after some numerical experiments, we observed that for the values of
$n$ and $q$ that we aim at using, the computational
  burden of the approach designed by \cite{rothman:2010} is too high
  for addressing our datasets, contrary to ours. Hence, in the following, we shall
not consider the method of \cite{rothman:2010} anymore. 

\subsubsection{Model selection issue}\label{sec:model_selection}

We can see that the estimator defined in (\ref{eq:lasso})
depends on a parameter $\lambda$ which tunes the sparsity level in $\widehat{\mathcal{B}}$. We propose to mix two standard
approaches to estimate the positions of the non null components in $\mathcal{B}$. 

We first use the  10-fold cross-validation method to choose
the $\lambda$ denoted $\lambda_{\textrm{CV}}$ minimizing the
cross-validation criterion.
This $\lambda$ is then used in the stability selection approach of
\cite{meinshausen:buhlmann:2010} which guarantees the robustness of the selected variables. This latter approach
can be described as follows.
The vector of observations $\mathcal{Y}$ is randomly split into
several subsamples of size $nq/2$ which is possible
  thanks to the whitening step.
For each subsample, the LASSO criterion is applied with $\lambda=\lambda_{\textrm{CV}}$
and the indices $i$ of the non null $\widehat{\mathcal{B}}_i$ are stored. 
Then, for a given threshold, we keep in the final
set of selected variables only the variables appearing a number of times larger than this threshold. 
In practice, we generated $5000$ subsamples of
$\mathcal{Y}$. 

Concerning the choice of the final threshold: we
propose either to take the one leading to the largest $p$-value of the
whitening test described in (\ref{eq:stat_test_2}) or the threshold 1. As we
shall see in Section \ref{sec:num_exp}, with
the first choice, mostly all the positions of the non null variables
in $\mathcal{B}$ are retrieved with some false positive. With the
second choice, all the true positions are
not recovered but there are no false positive. Moreover, the second choice guarantees a stability of the selected variables since
only the variables which are chosen at each of the 5000 subsamplings of the data are finally kept.


 \section{Simulation study}\label{sec:num_exp}

The goal of this section is to assess the statistical performance of
our methodology. In order to emphasize the benefits of using a
whitening approach from the variable selection point of view, we shall
first compare our approach to standard methodologies. Then, we shall analyze the performance
of our statistical test for choosing the best dependence
modeling. Finally, we shall investigate the performance of our model
selection criterion.

To assess the performance of
these different methodologies, we generated observations
$\boldsymbol{Y}$ according to Model (\ref{eq:model:matriciel}) with $q=1000$,
$p=3$, $n=30$ and different dependence modelings, namely different
matrices $\boldsymbol{\Sigma}_q$ corresponding to the AR(1) model
described in (\ref{eq:AR1}) with $\sigma=1$ and $\phi_1=0.7$ or 0.9.

Note that we have chosen the values of the parameters $p$, $q$ and $n$
in order to be as close as possible to the real data that we plan to analyze in
Section \ref{sec:real}.

We shall also investigate the effect of the sparsity and of the signal
to noise ratio. In the following, the sparsity level 
corresponds to the number of non null elements in $\mathcal{B}$
divided by the total number $nq$ of elements of $\mathcal{B}$. Different signal to noise ratios are
obtained by multiplying $\boldsymbol{B}$ in (\ref{eq:model:matriciel})
by a coefficient $\kappa$.

\subsection{Variable selection performance}

The goal of this section is to compare
the performance of our different whitening strategies presented above to standard existing methodologies.
More precisely, we shall compare our approaches to the classical
ANOVA method (denoted \textsf{ANOVA}), the standard Lasso
(denoted \textsf{Lasso}), namely the Lasso approach without
the whitening step and to \textsf{sPLSDA} devised by \cite{LeCao2011} and implemented in the \textsf{mixOmics} R package, which is widely used in the metabolomics field. 
By \textsf{ANOVA}, we mean the classical one-way ANOVA applied to each column of the observations matrix $\boldsymbol{Y}$
without taking the dependence into account. 

In the following, the different whitening
approaches that we propose will be denoted by \textsf{AR1} and
\textsf{Nonparam}. They are described in Sections
\ref{subsec:param} and \ref{subsec:nonparam}, respectively. These methods will also be
compared to the \textsf{Oracle} approach which assumes that the matrix
$\boldsymbol{\Sigma}_q$ is known, which is never the case in practical
situations. 

For comparing these different methods, we shall use two classical
criteria: ROC curves and AUC (Area Under the ROC Curve). ROC curves display the true positive rates as
a function of the false positive rates and the closer to one the AUC
the better the methodology. Since \textsf{sPLSDA} only selects relevant metabolites but does not assign them to a level
of the factor, we shall consider that as soon as a relevant metabolite is selected it is a true positive which gives an advantage
to \textsf{sPLSDA}.

\begin{figure}[t!]
\hspace{-2mm}
\begin{tabular}{@{}l@{}c@{}c@{}c@{}c@{}c@{}r@{}}
  & \multicolumn{2}{c}{$s=0.01$} & \hspace{1em} & \multicolumn{2}{c}{$s=0.3$} & \\
  \cline{2-3}\cline{5-6}
  & $\phi_1=0.7$ & $\phi_1=0.9$ & &$\phi_1=0.7$ & $\phi_1=0.9$ & \\
  \rotatebox{90}{\hspace{6em}\small TPR} & 
\includegraphics[scale=0.235]{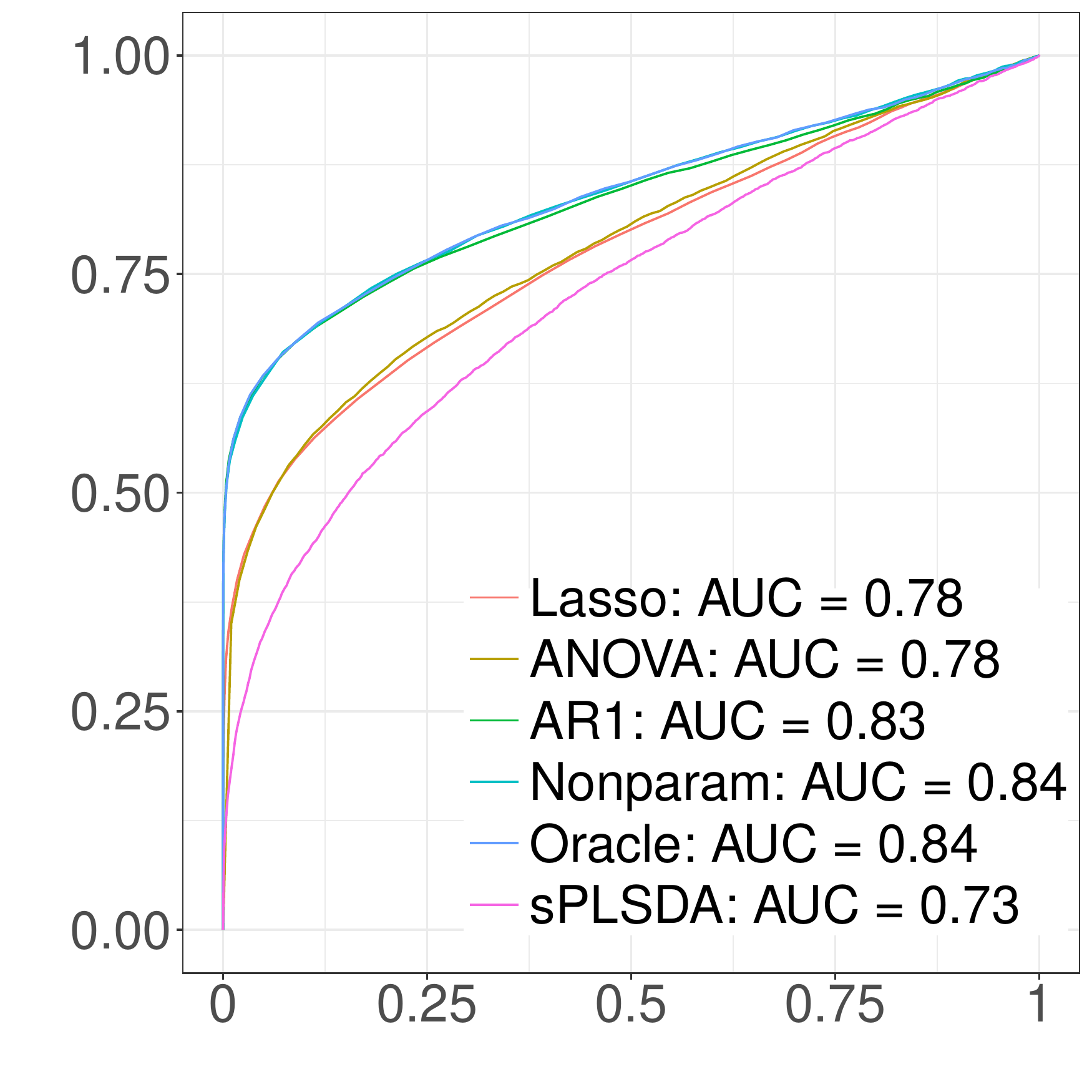}
& \includegraphics[scale=0.235]{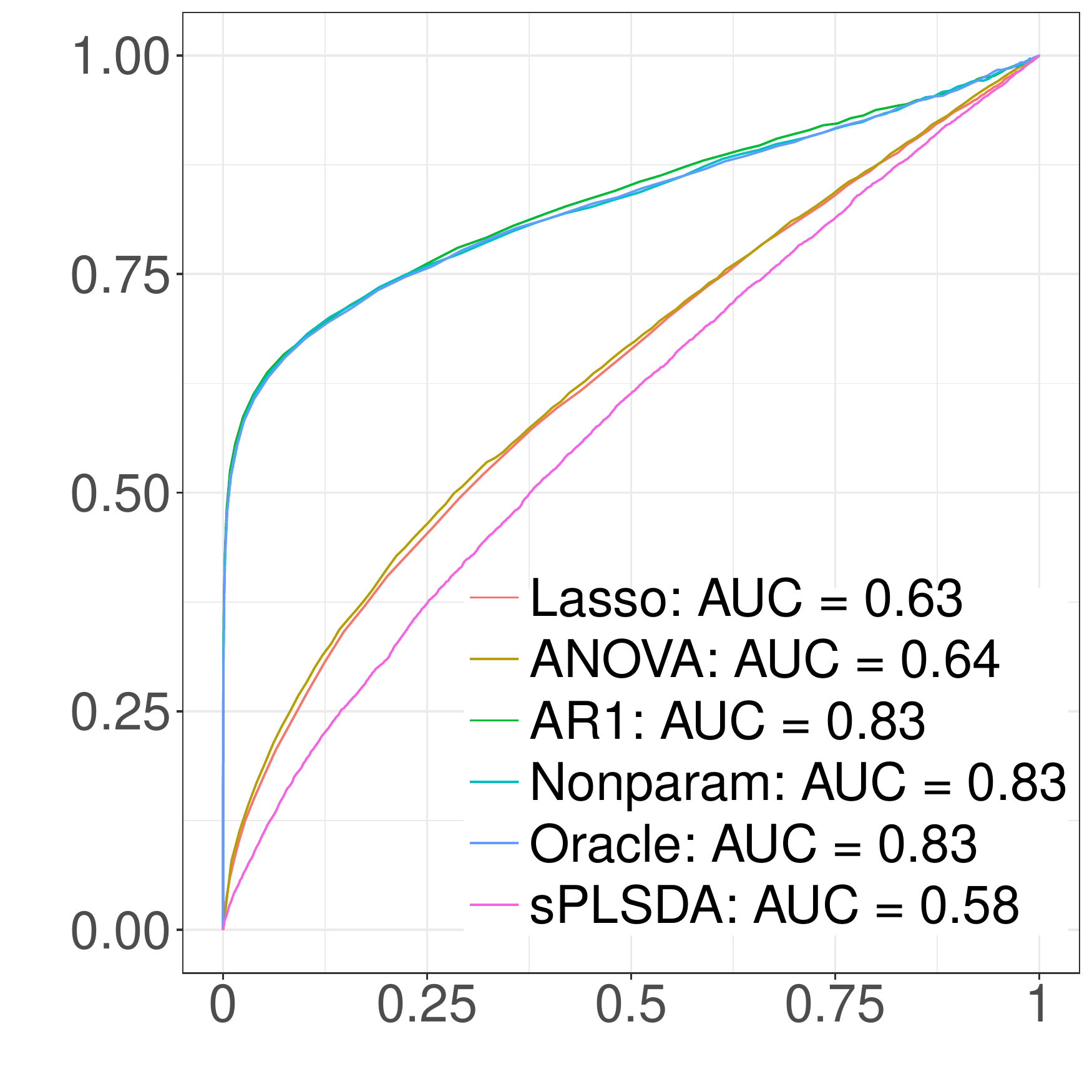}
& & \includegraphics[scale=0.235]{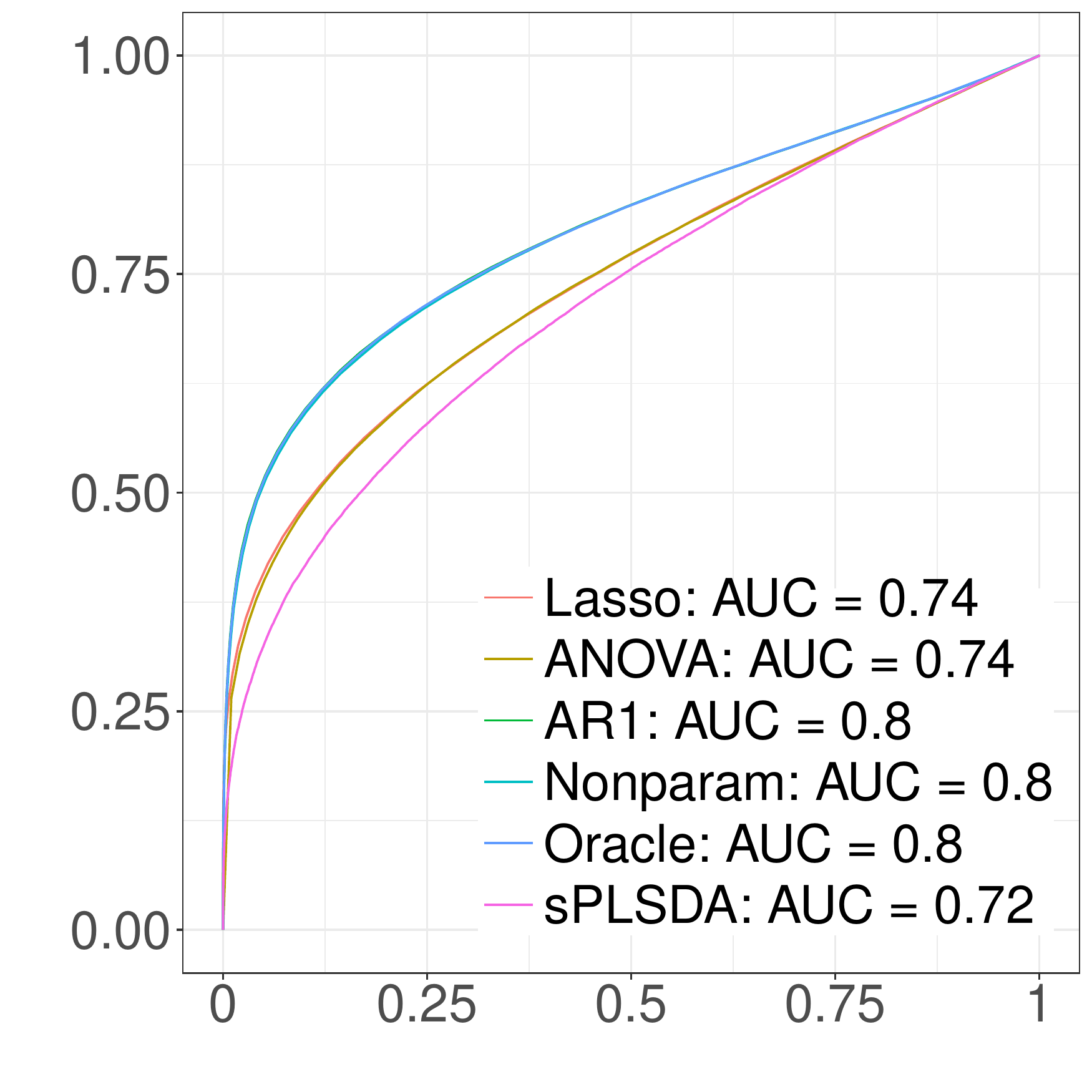}
& \includegraphics[scale=0.235]{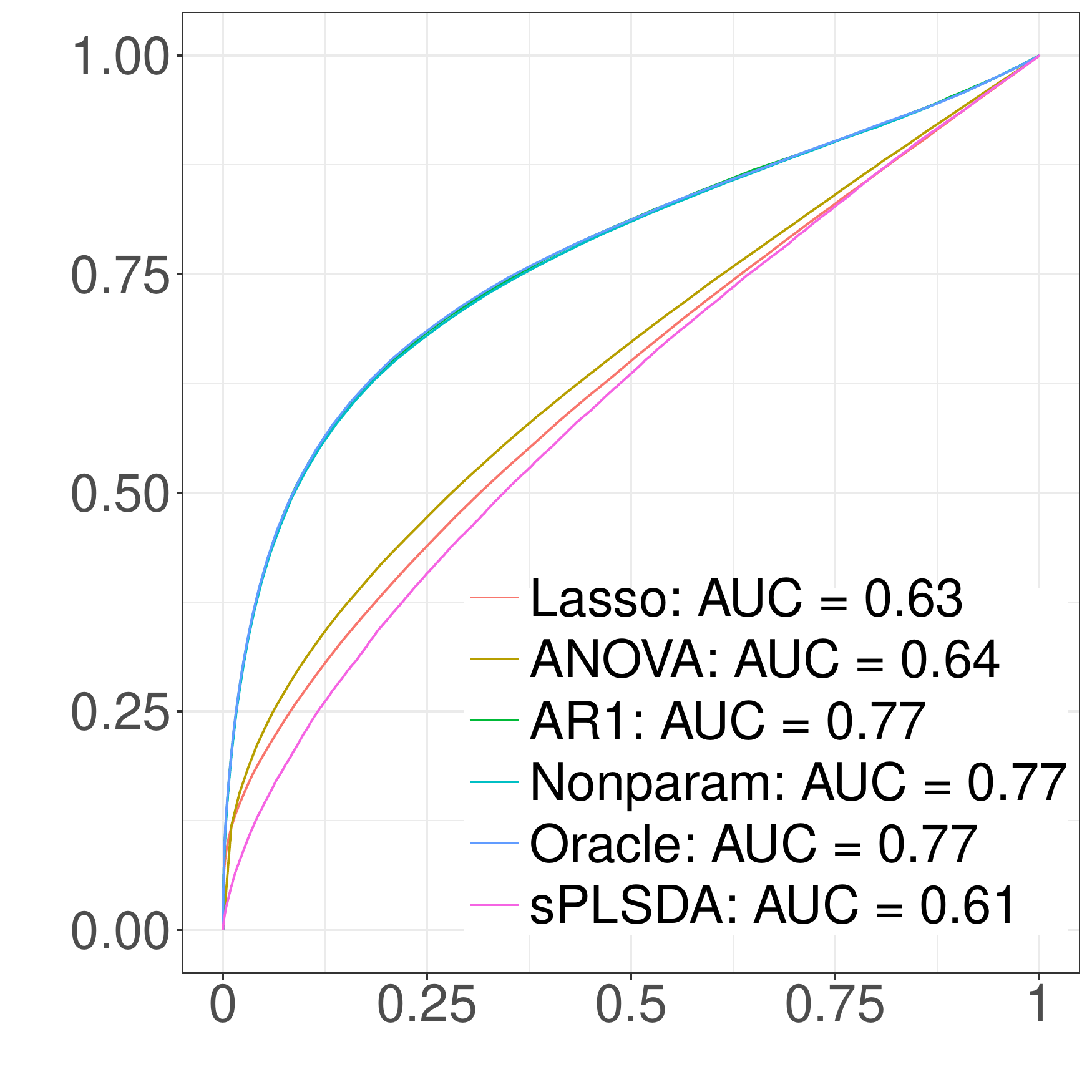} & \rotatebox{90}{\hspace{6em}\small $\kappa=1$}\\
\rotatebox{90}{\hspace{6em}\small TPR} & \includegraphics[scale=0.24]{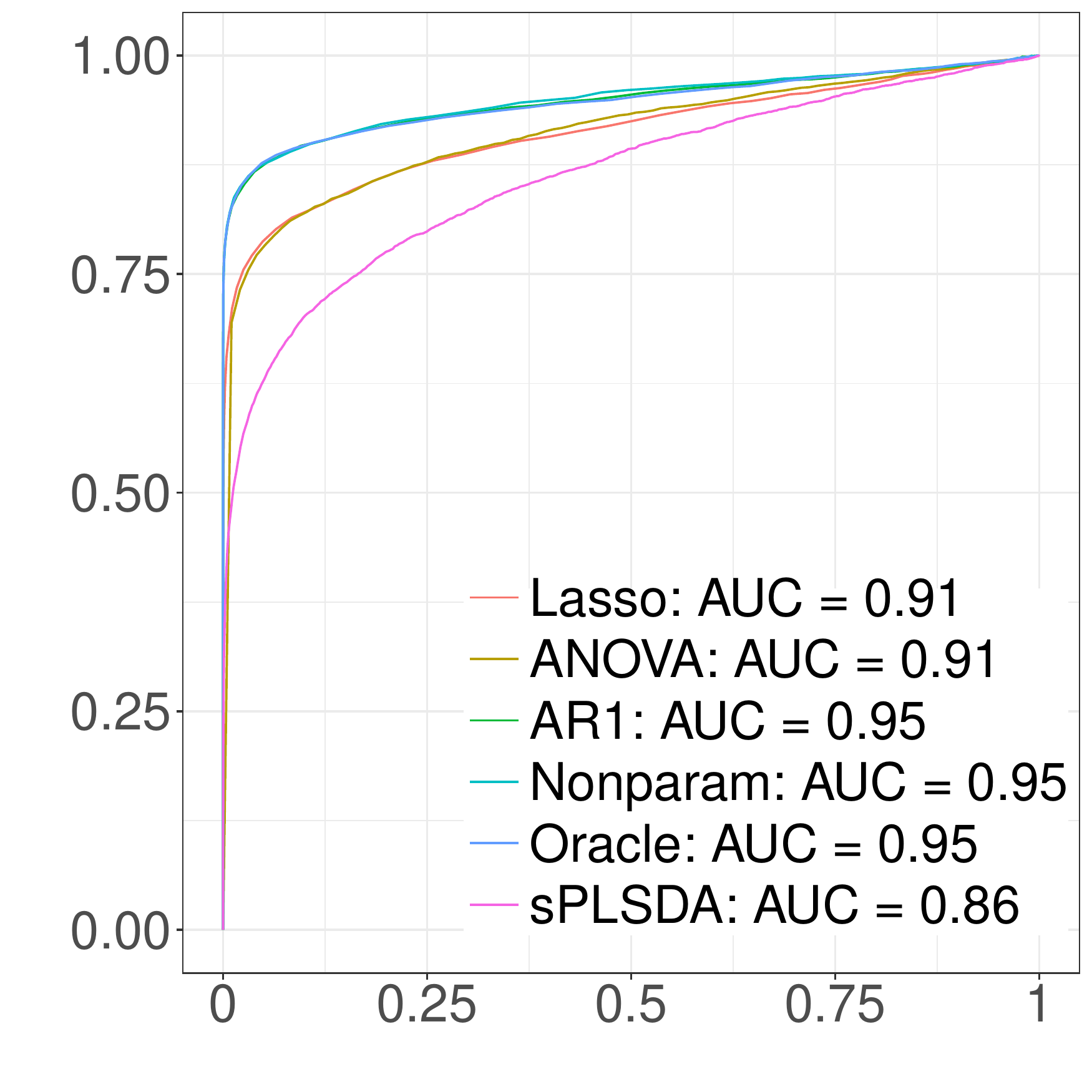}
& \includegraphics[scale=0.235]{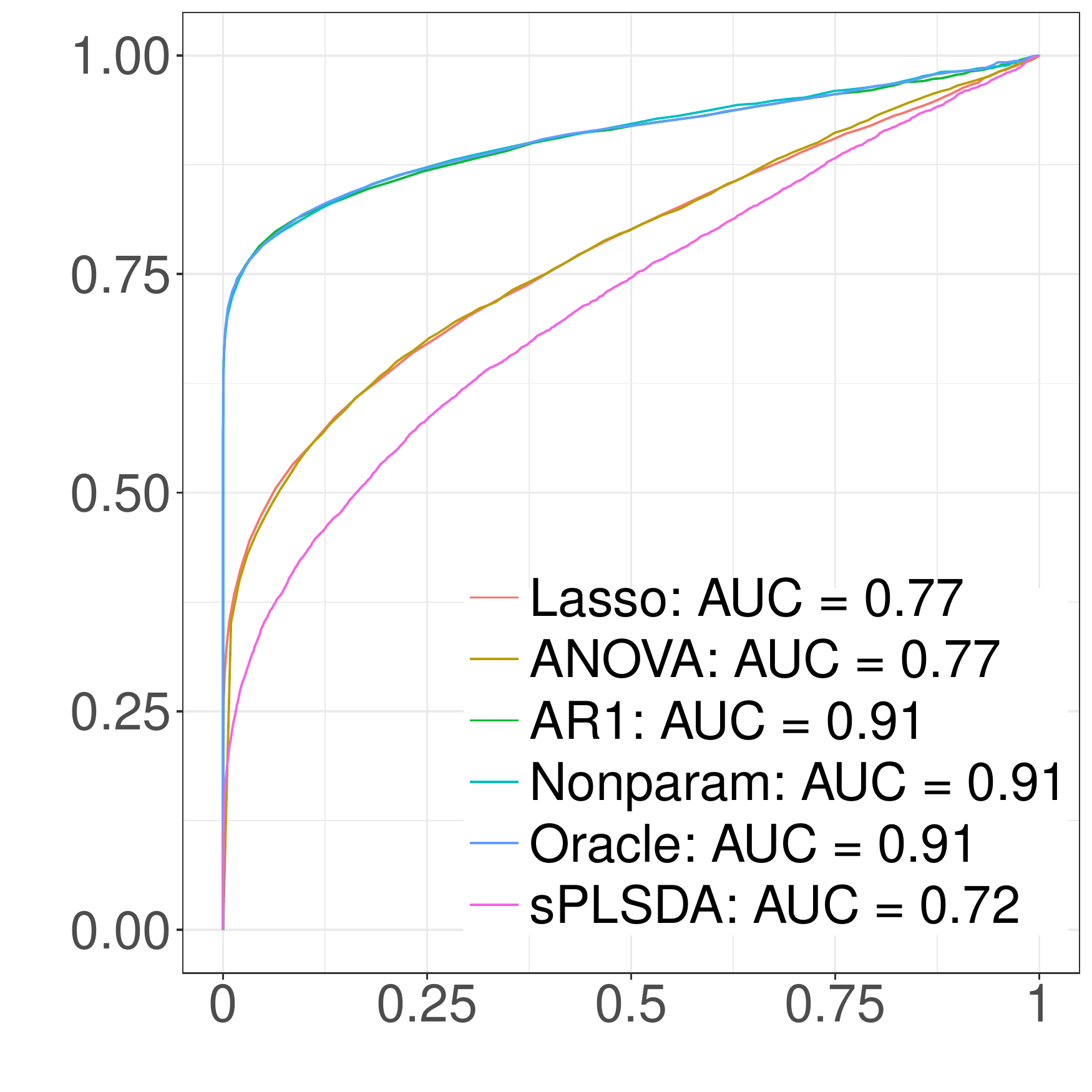}
& & \includegraphics[scale=0.235]{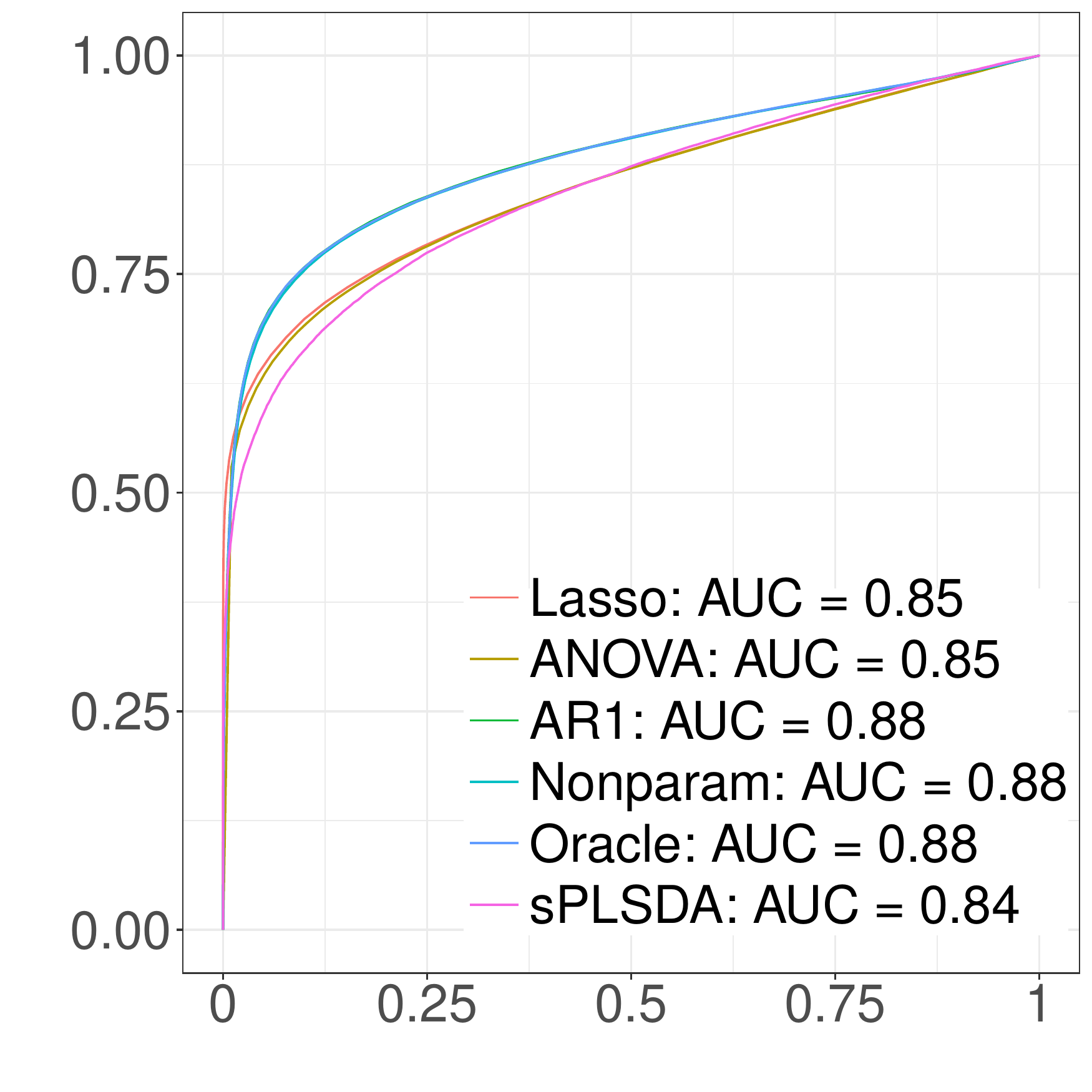}
& \includegraphics[scale=0.235]{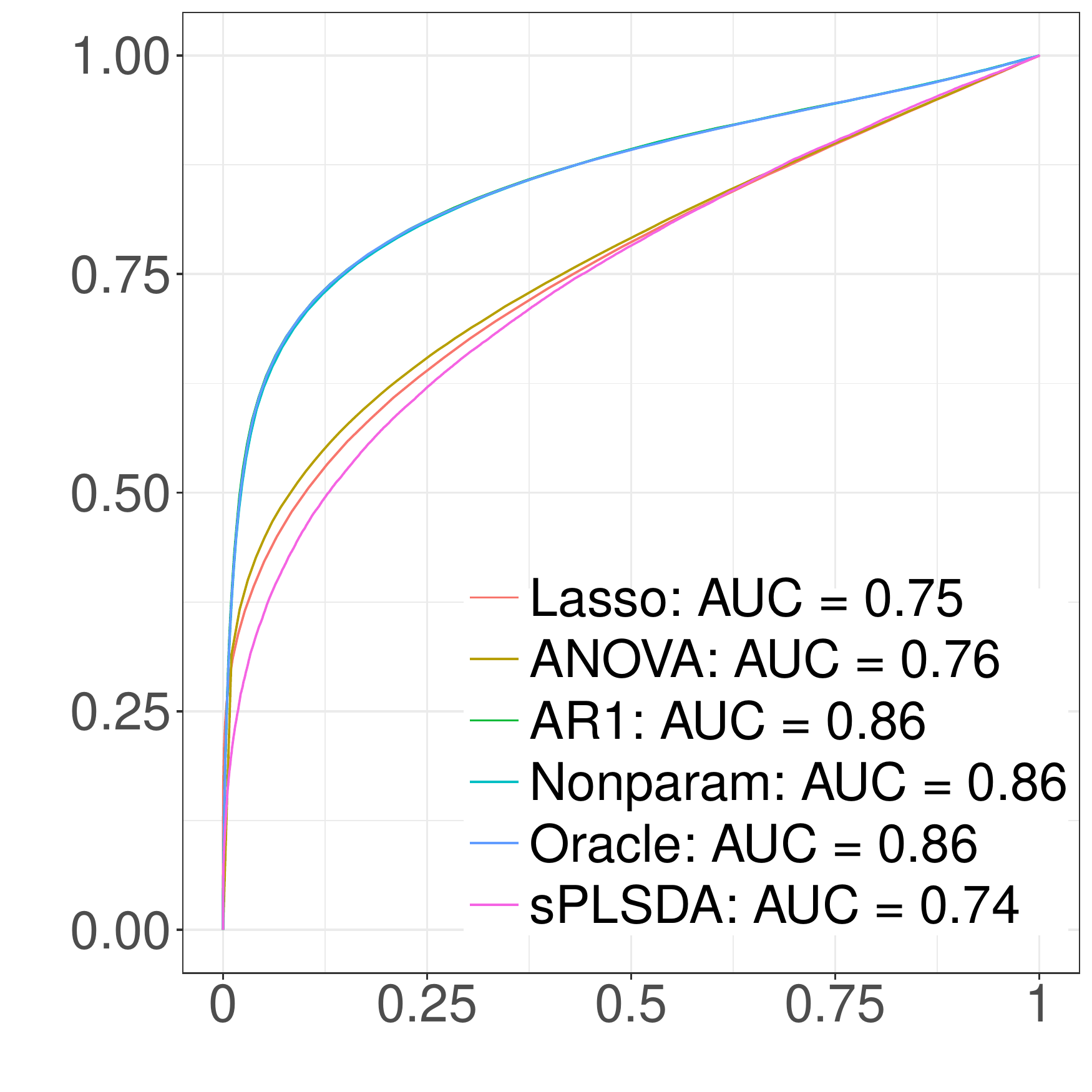} & \rotatebox{90}{\hspace{6em}\small $\kappa=2$}\\
& \multicolumn{4}{c}{FPR} & \\
\end{tabular}
\caption{Means of  the ROC curves  obtained from 200  replications for
  the  different  methodologies  in  the  AR(1)  dependence  modeling;
  $\kappa$ is linked to the signal to noise ratio (first row: $\kappa=1$, second row $\kappa=2$); $\phi_1$ is the
  correlation level in the AR(1) and  $s$ the sparsity level (i.e. the
  fraction of nonzero elements) in the vector of true parameters.\label{fig:AR1_1}}
\end{figure}

We can see  from Figure \ref{fig:AR1_1} that in the  case of an AR(1)
dependence,  taking  into  account  this  dependence  provides  better
results  than \textsf{sPLSDA} and than  approaches that consider  the  columns of  the  residual  matrix  as
independent.  Moreover,  we  observe  that  the
performance of the non parametric modeling  are on a par with those of
the parametric and the oracle ones.   We also note that the larger the
sparsity level the  smaller the difference of  performance between the
different approaches.  However,  the larger the signal  to noise ratio
the better the performance of the different methodologies.

\subsection{Choice of the dependence modeling}

\begin{figure}[!h]
\centering
\includegraphics[scale=0.4]{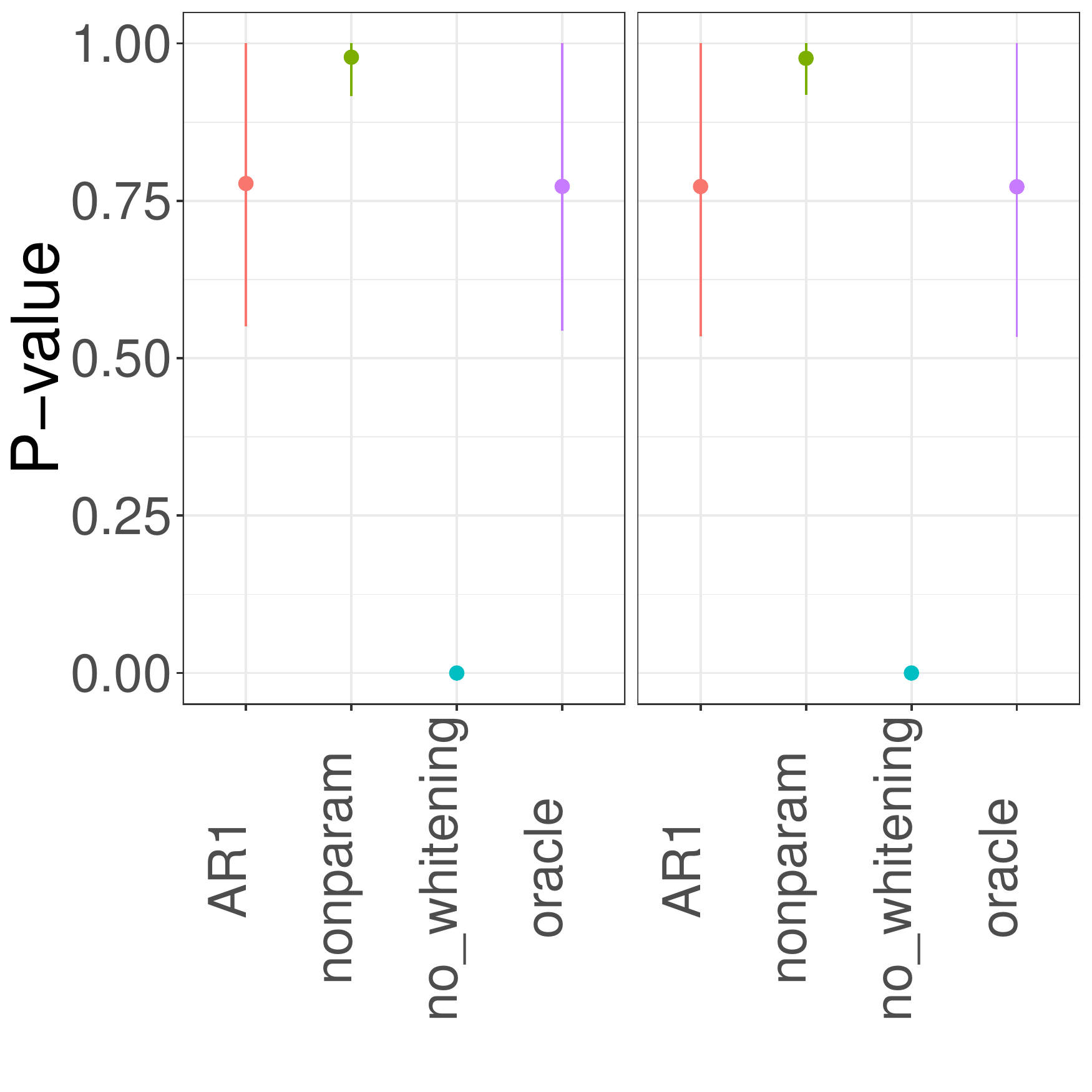}
\caption{Means and standard deviations of the $p$-values of the test
   described in Section \ref{sec:whitening_test} for the different
   approaches in the AR(1) dependence modeling when $\phi_1=0.7$ (left)
   and $\phi_1=0.9$ (right).\label{fig:test}}
\end{figure}

The goal of this section is to assess the performance of the dependence modeling
strategy that we proposed in Section \ref{sec:whitening_test}. We
generated observations $\boldsymbol{Y}$ with the parameters described
at the beginning of Section \ref{sec:num_exp} in the case of an AR(1)
dependence, for a sparsity level of 0.01 and when $\kappa=1$. The corresponding results are displayed
in Figure \ref{fig:test}.

We observe from this figure that our test provides
$p$-values close to zero in the case where no whitening strategy is
used (\textsf{Lasso}) and that when one of the proposed whitening
approaches is used the $p$-values are larger than 0.7.

\subsection{Choice of the model selection criterion}

We investigate hereafter the performance of our model selection
criterion described in Section \ref{sec:model_selection}.

\begin{figure}[htbp!]
\centering
\includegraphics[scale=0.25]{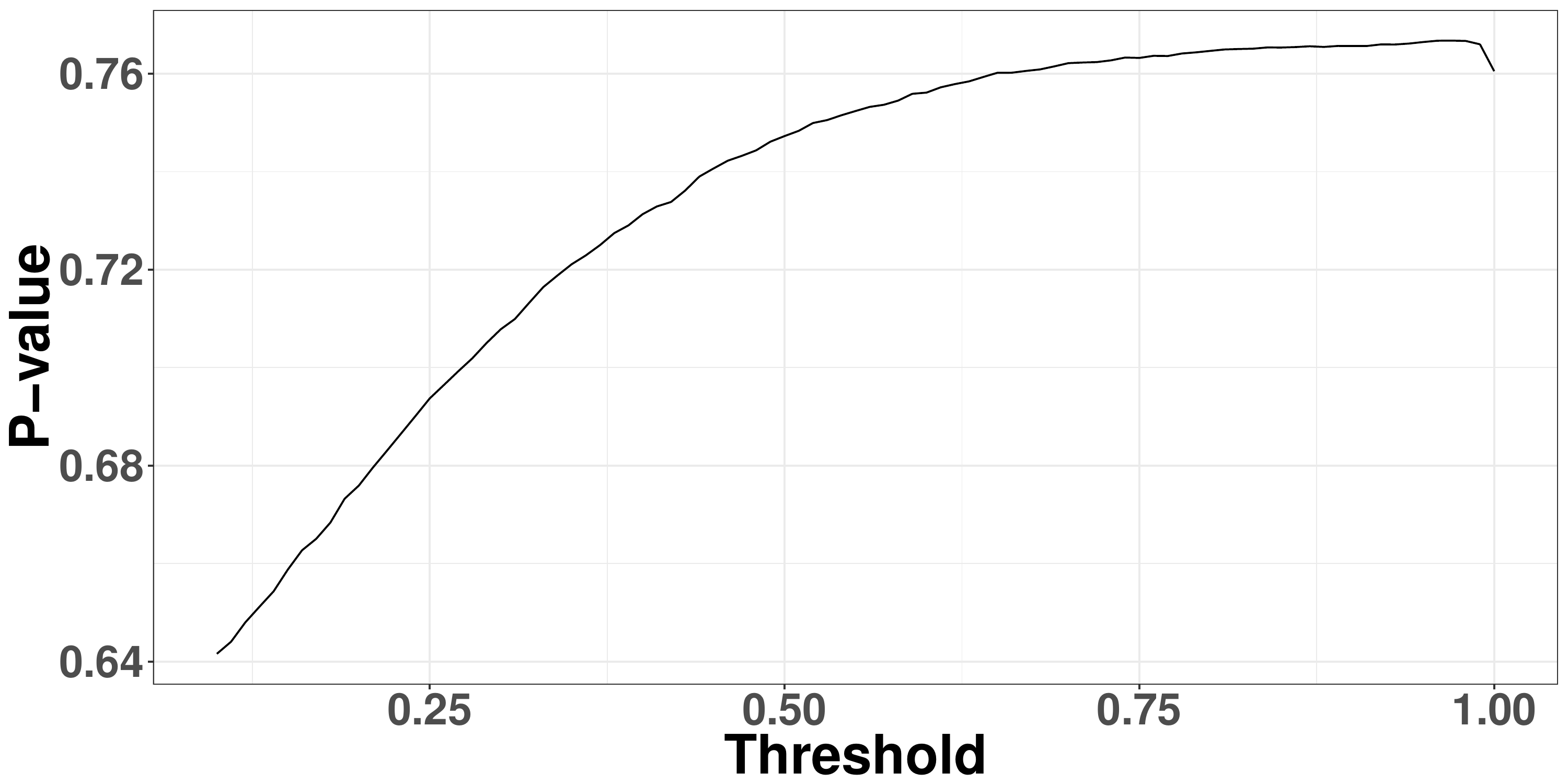}
\caption{Means of the $p$-values associated to the test statistic defined in (\ref{eq:stat_test_2}) obtained
  from 5000 replications when $\kappa=1$. 
\label{fig:pval_simul}}
\end{figure}

Figure \ref{fig:pval_simul} displays the means of the $p$-values of
the test described in \ref{sec:whitening_test} obtained from 5000 replications of the observations $\boldsymbol{Y}$
generated with the parameters described
at the beginning of Section \ref{sec:num_exp} in the case of an AR(1)
dependence with $\phi_1=0.9$ and $\kappa=1$. 

We observe
from this figure that the $p$-values are all the more high that the
thresholds are large. 

\begin{figure}[!htbp!]
\centering
\includegraphics[scale=0.2]{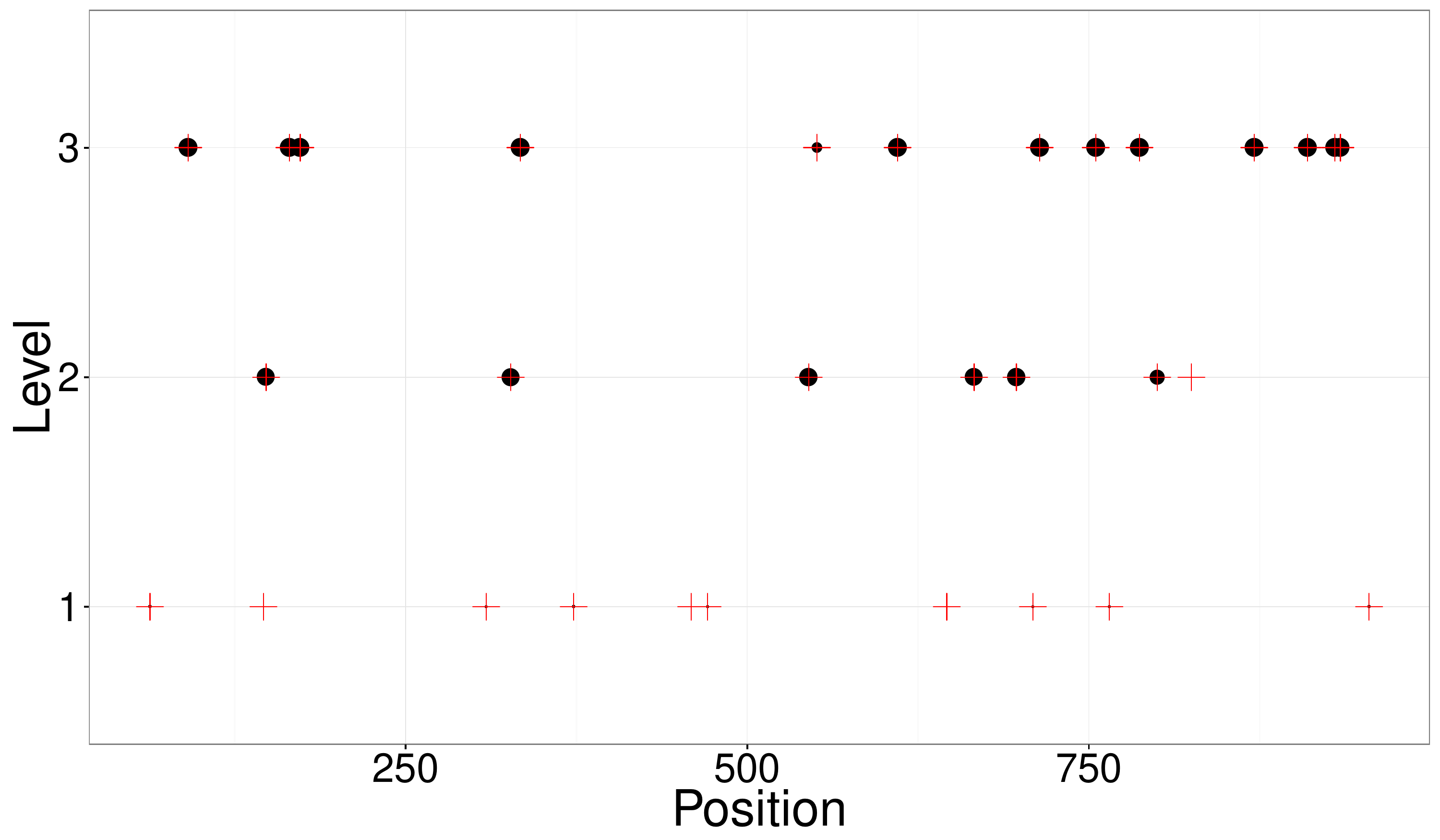}
\includegraphics[scale=0.2]{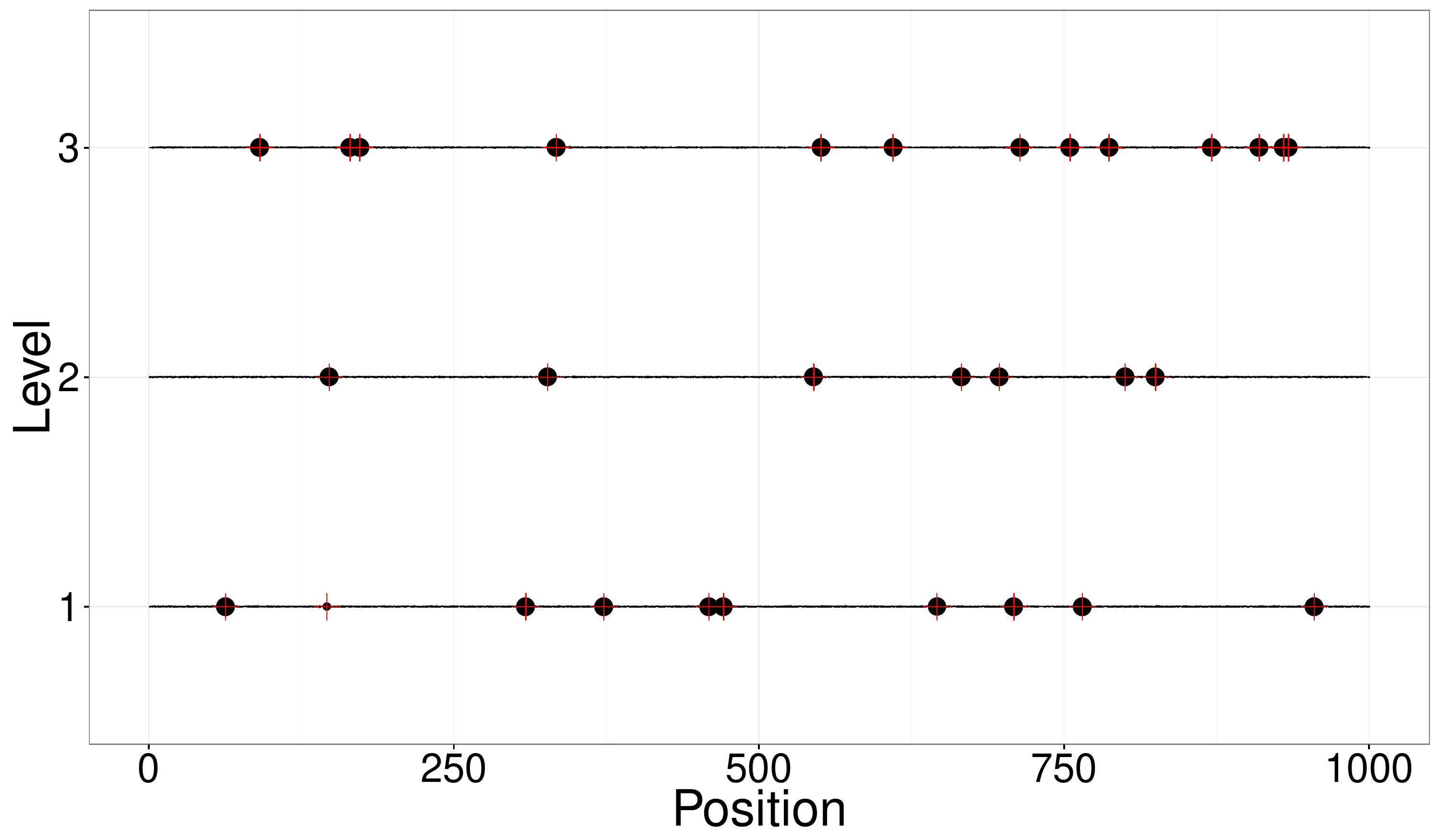}
\includegraphics[scale=0.2]{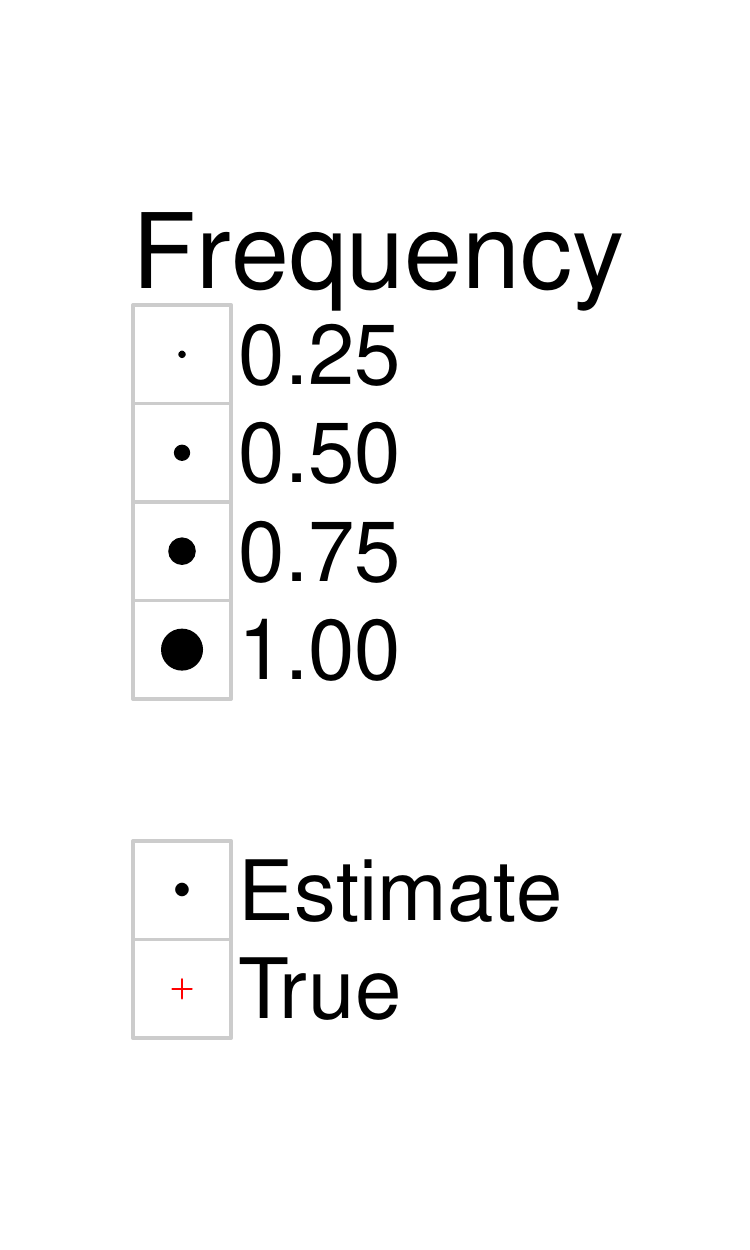}
\caption{Positions of the variables selected by our
  approach  ('$\bullet$') when  $\kappa=10$.  Values  on the  $y$-axis
  correspond to the 3 levels
  of the factor. The results obtained when the threshold
  is equal to 1 are on the left and the results when the threshold is
  obtained by maximizing the $p$-value are on the right. The size
  of the bullets are all the more large that 
  the number of times where a variable has been selected is large.
\label{fig:boulier_simuls}}
\end{figure}

Figure   \ref{fig:boulier_simuls}  displays   with  bullets
('$\bullet$')  the   positions  of  the  variables   selected  by  our
three-step approach  for the two  possible choices of  thresholds from
500 replications of $\boldsymbol{Y}$  obtained  with   the  parameters
described at the beginning of Section \ref{sec:num_exp} in the case of
an  AR(1) dependence  with $\phi_1=0.9$  and $\kappa=10$.

   We observe
from  this figure  that  mostly  all the  positions  of  the non  null
variables in $\mathcal{B}$ are retrieved with some false positive when
the  threshold  is obtained  by  maximizing  the $p$-value.  When the
threshold is equal to 1, there are  no false positive but all the true
positions are not recovered.

\subsection{Numerical performance}

In order to investigate the computational burden of our approach, we generated matrices $\boldsymbol{Y}$
satisfying Model (\ref{eq:model:matriciel}) with $n=30$ and $q\in\{100,200,\dots,1000\}$. Here, the rows of the matrix
$\boldsymbol{E}$ are generated as realizations of an AR(1) process and the level of sparsity $s$ 
of $\boldsymbol{B}$ is equal to 0.01.

\begin{figure}
\centering
\includegraphics[scale=0.8]{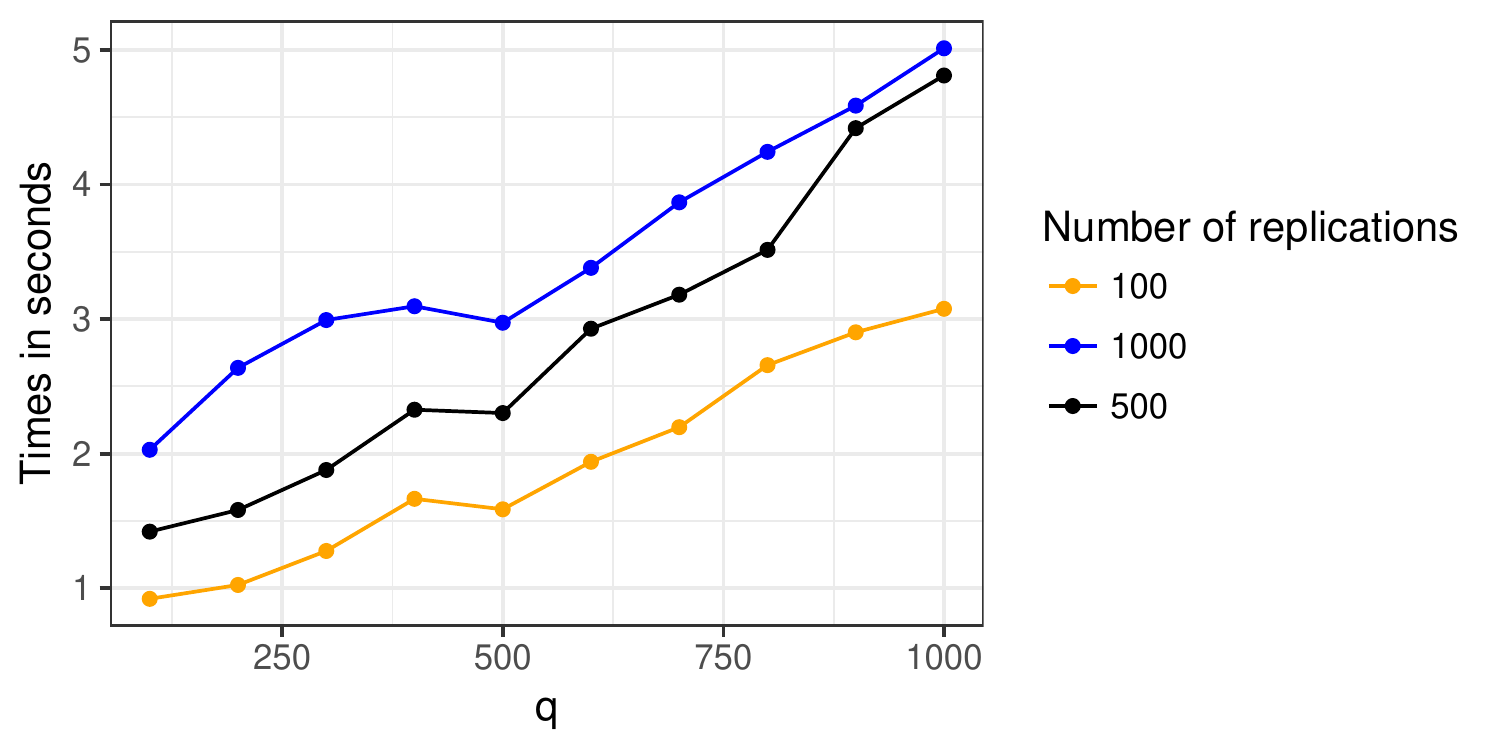}
\caption{Computational times (in seconds) of \textsf{MultiVarSel}.\label{fig:time}}
\end{figure}

Figure \ref{fig:time} displays the computational times of \textsf{MultiVarSel}
obtained from a computer having the following configuration: RAM 16 GB, CPU $8\times 3.6$ GHz for different 
number of replications in the stability selection stage.
We can see from this figure that the computational burden  of \textsf{MultiVarSel} is very low and that it takes
only a few seconds to analyze matrices having 1000 columns.

\section{Application to a LC-MS dataset}\label{sec:real}

In this section, our three-step methodology implemented in the R package \textsf{MultiVarSel} and available from the CRAN, 
is applied to a LC-MS (Liquid Chromatography-Mass Spectrometry) data set made of African copals samples. 
The samples correspond to  ethanolic extracts of copals produced by trees belonging to two genera \textit{Copaifera} (C) and \textit{Trachylobium} (T) with a second level of classification coming 
from the geographical provenance of the \textit{Copaifera} samples (West (W) or East (E) Africa). Since all the \textit{Trachylobium} samples come from East Africa, we have a single factor
having three levels: CE, CW and TE such that $n_{\textrm{CE}}=9$, $n_{\textrm{CW}}=8$ and $n_{\textrm{TE}}=13$.

In this section, we also compare the performance of our method with those of other techniques which are widely used in metabolomics.

\subsection{Data pre-processing}

LC-MS chromatograms were aligned using the R package
XCMS proposed by \cite{Smith2006} with the following parameters: a signal to noise
ratio threshold of 10:1 for peak selection,  a step size of 0.2 min
and a minimum difference in \textit{m}/\textit{z} for peaks with
overlapping retention times of 0.05 amu. 
Sample filtering was also performed: To be considered as informative, as suggested by \cite{Kirwan2013}, a peak was required to be present in at least 80\% of the samples. 
Missing values imputation was realized using the KNN algorithm described in \cite{Hrydziuszko2012}.
Subsequently, the spectra were normalized to equalize signal
intensities to the median profile in order to reduce any variance
arising from differing dilutions of the biological extracts and
probabilistic quotient normalization (PQN) was used, see
\cite{Dieterle2006} for further details. In order to reduce the size of the data matrix,
selection of the adducts of interest [M+H]$^{+}$ was then performed 
using the CAMERA package of \cite{Kuhl2012}. A $n\times q$ matrix $\boldsymbol{Y}$ was then obtained and submitted to the statistical analyses.

\subsection{Application of our three-step approach}

The observations matrix $\boldsymbol{Y}$ is first centered and scaled in order to ensure that the
empirical mean in each column is 0 and that the empirical variance is 1.

\subsubsection{First step} A one-way ANOVA is fitted to each column of the observation matrix 
 $\boldsymbol{Y}$ in order to have access to an
  estimation $\widehat{\boldsymbol{E}}$ of the residual matrix
  $\boldsymbol{E}$. Then, the test proposed in Section \ref{sec:whitening_test} is applied. 
We found a $p$-value equal to zero
which indicates that the columns of $\widehat{\boldsymbol{E}}$ cannot
be considered as independent and hence that applying the whitening
strategy should improve the results.

\subsubsection{Second step} The different whitening strategies described in Section \ref{sec:estim_sigma_q} were applied and the highest 
$p$-value for the test described in Section \ref{sec:whitening_test} is obtained for the nonparametric whitening. 
More precisely, the $p$-values obtained for the AR(1) and the nonparametric dependence modeling
are equal to $1.5\times 10^{-4}$ and 0.5107, respectively. Hence, in the following we shall use the nonparametric modeling.

\subsubsection{Third step} The Lasso approach described in Section \ref{sec:lasso} was then applied to the
  whitened observations where  $\widehat{\boldsymbol{\Sigma}}_q$ is obtained by using the
  nonparametric modeling. The stability selection is then used with 5000 replications and a threshold equal to 1 in order to avoid false positive.

\begin{figure}[!h]
\begin{center}
\includegraphics[scale=0.15]{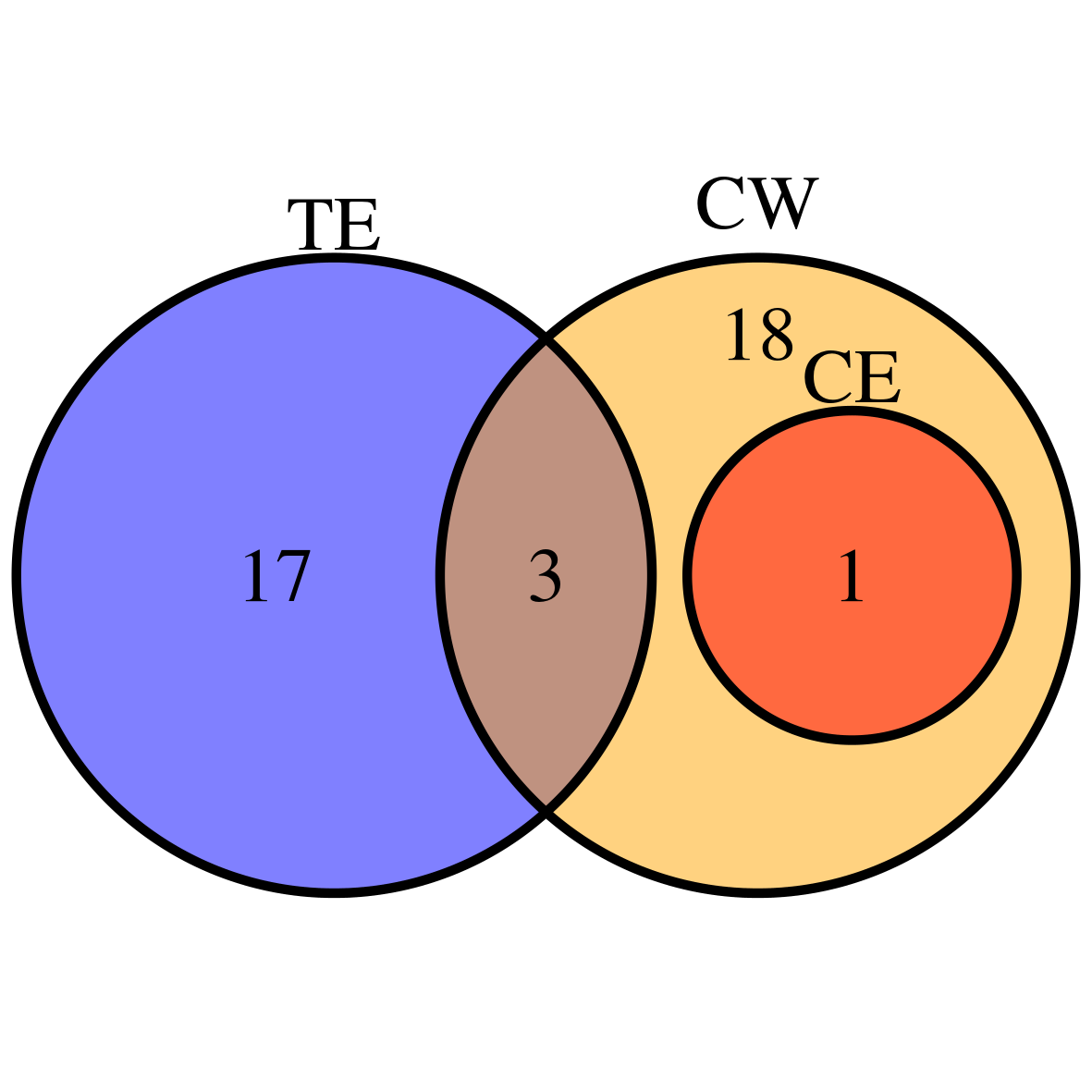}
\vspace{-10mm}
\caption{Venn diagram of the metabolites selected for each class by  \textsf{MultiVarSel}
 using a threshold equal to 1 in the stability selection stage.\label{fig:real_boulier}}
\end{center}
\end{figure}

The Venn diagram of Figure \ref{fig:real_boulier} displays the repartition of the selected metabolites among the different classes CE, TE and CW.
We can see from this figure that at least one metabolite
is selected as a marker for each class (20 for TE, 22 for CW and 1 for
CE) for a total of 39 unique metabolites. More precisely, our methodology leads to a list of metabolites that mainly characterize a single class.

\subsection{Comparison with existing methods}

The goal of this section is to compare the performance of our approach with those of methodologies classically used in metabolomics 
such as partial least square discriminant analysis (PLS-DA) 
and sparse partial least square discriminant analysis (sPLS-DA) devised by \cite{LeCao2011} and implemented in the R package \verb|MixOmics|. 

\begin{figure}[!h]
\centering
\begin{tabular}{cc}
\includegraphics[scale=0.7]{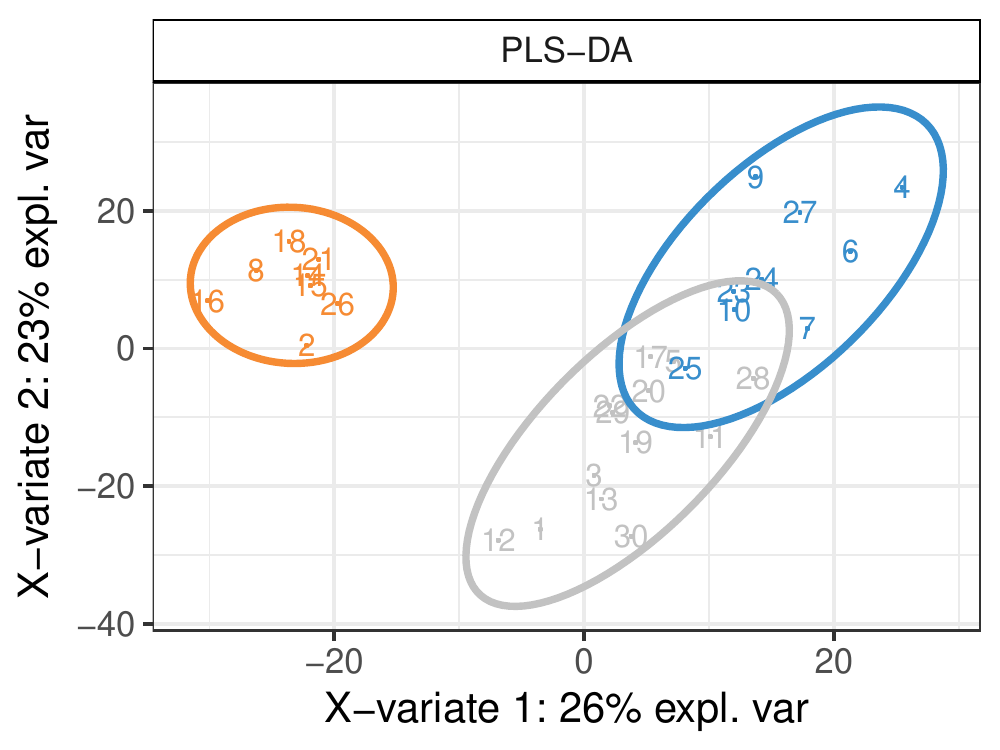} 
&\includegraphics[scale=0.55]{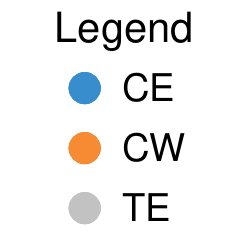} \\
\includegraphics[scale=0.7]{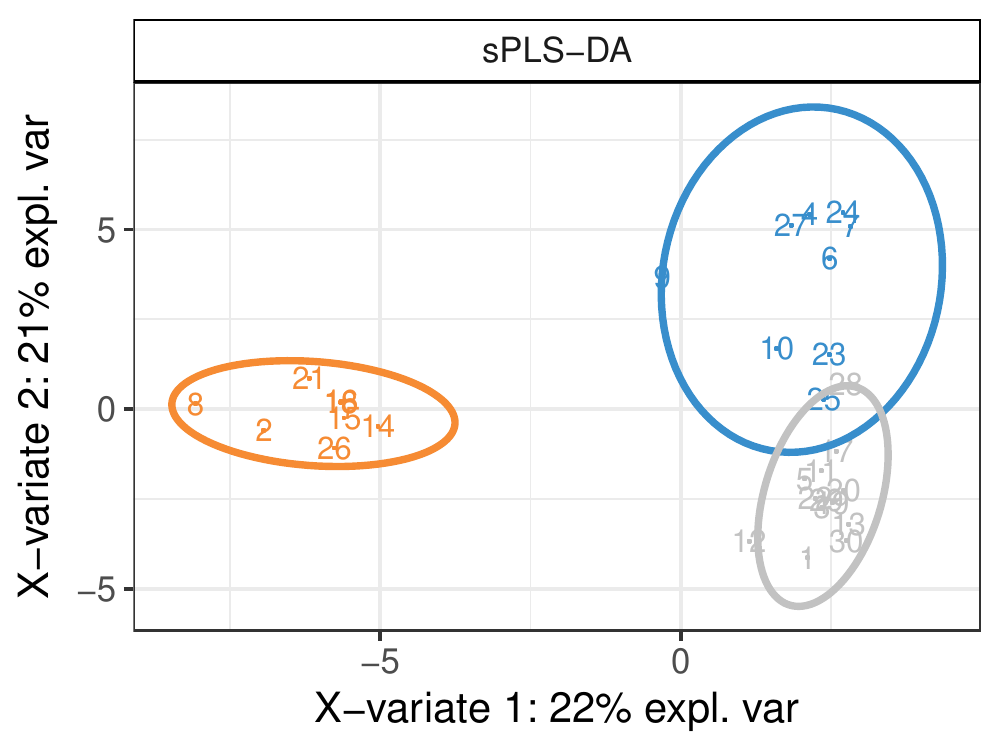} & \\
\end{tabular}
\caption{2D scores plot of the PLS-DA and the sPLS-DA.\label{fig:2dplsda}}
\end{figure}

As recommended by \cite{LeCao2011}, we used two components for PLS-DA and sPLS-DA. Moreover, in order to make sPLS-DA comparable with our approach, 
20 variables are kept for each component in the sPLS-DA methodology. The corresponding results  are
displayed in Figure \ref{fig:2dplsda}. We can see from this figure that sPLS-DA exhibits better classification performance than the standard PLS-DA.

\begin{figure}[!h]
\centering
\begin{tabular}{cc}
\includegraphics[scale=0.7]{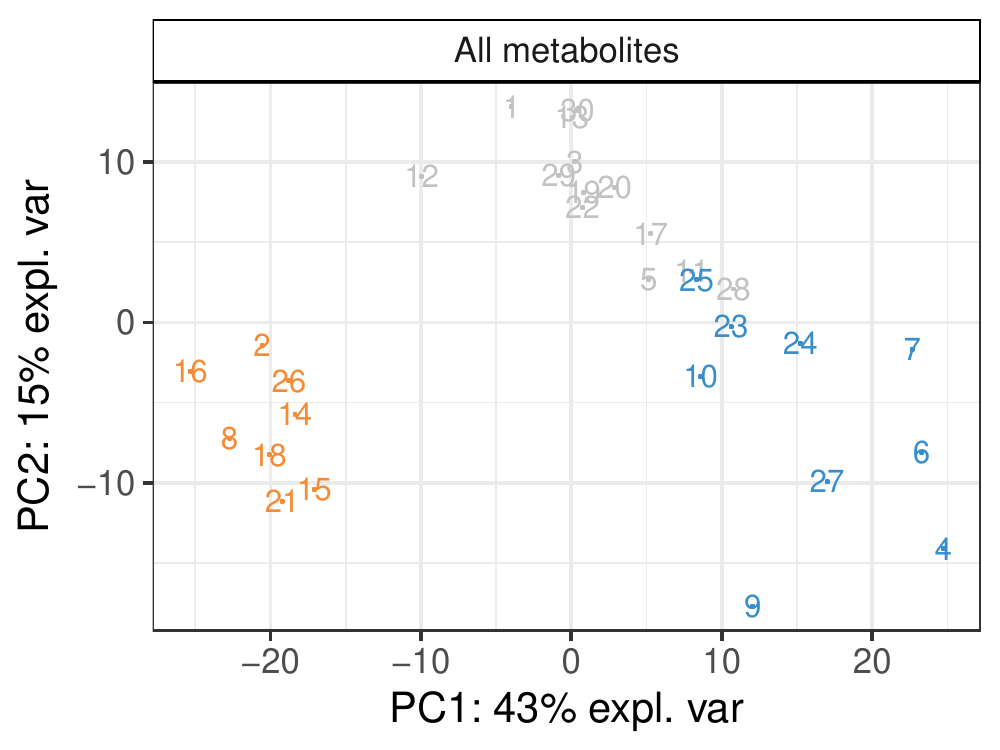} & 
\includegraphics[scale=0.55]{legende_plot2d_plsda.pdf}\\
\includegraphics[scale=0.7]{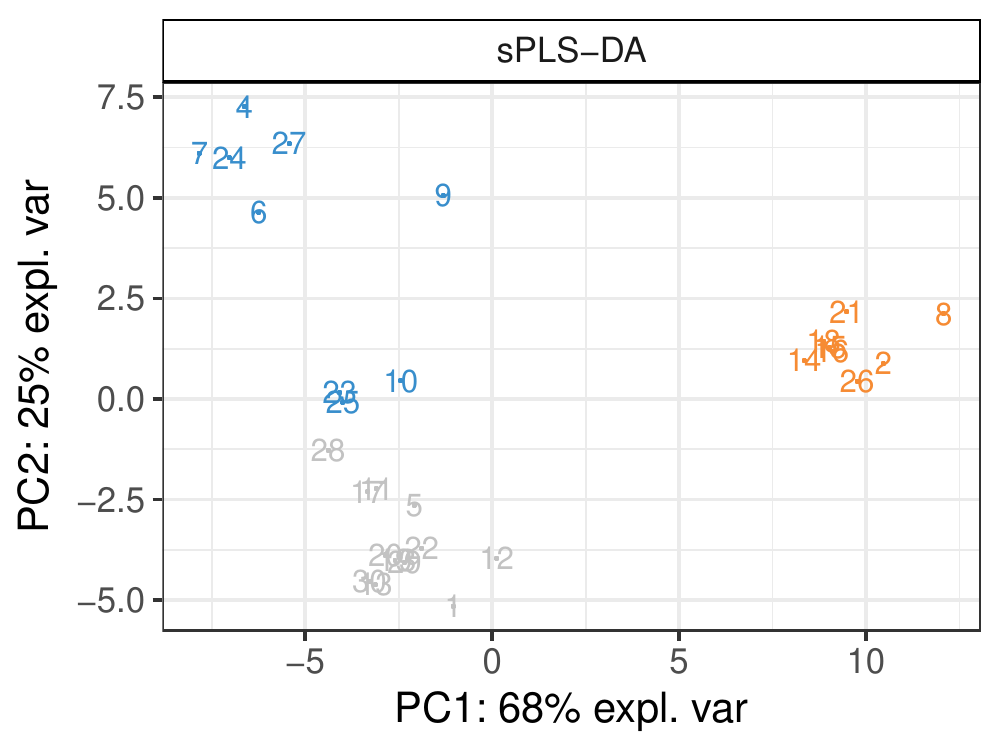}
& \\
\includegraphics[scale=0.7]{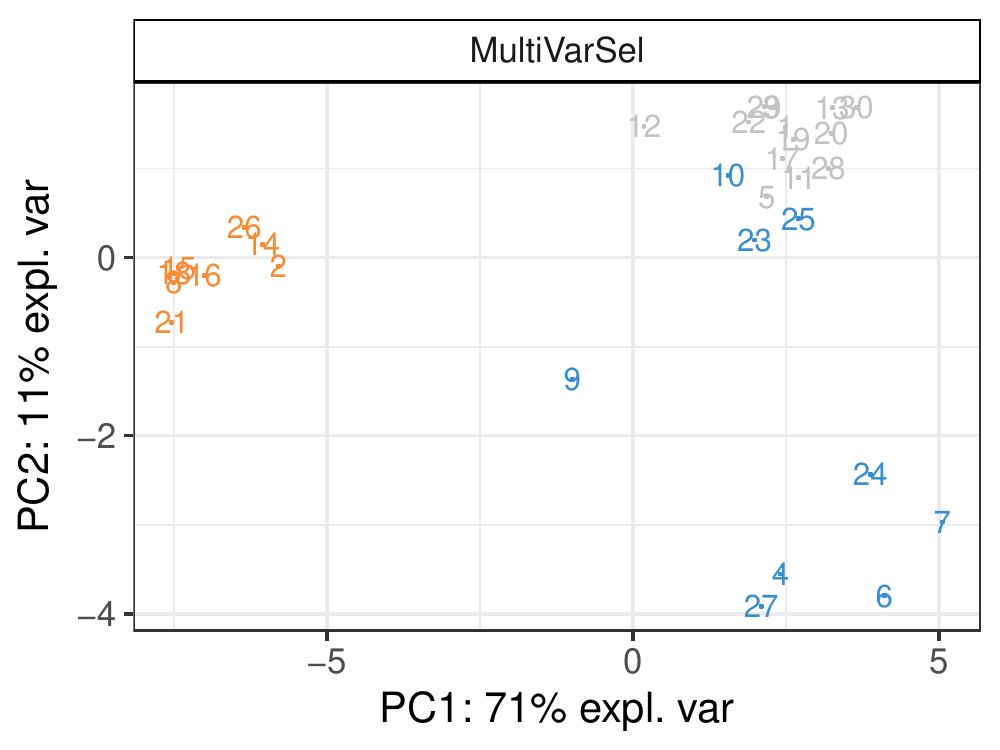}
& \\
\end{tabular}
\caption{PCA with all the metabolites and with the metabolites selected by sPLS-DA and our approach \textsf{MultiVarSel}.\label{fig:pcaall}}
\end{figure}

Since PLS-DA does not include a variable selection step we shall compare our approach only to sPLS-DA in the following. 
For comparing these methodologies Figure \ref{fig:pcaall} displays the PCA obtained when all the metabolites are kept on the one hand 
and when the metabolites are those selected by sPLS-DA or by our methodology on the other hand. We can see from this figure that, one the hand,
the approaches containing a variable selection step exhibit better classification performance and that, on the other hand, 
sPLS-DA and our method show similar performance from the classification point of view even if our approach is not designed for this purpose.

\begin{figure}[!h]
\begin{center}
\includegraphics[scale=0.75]{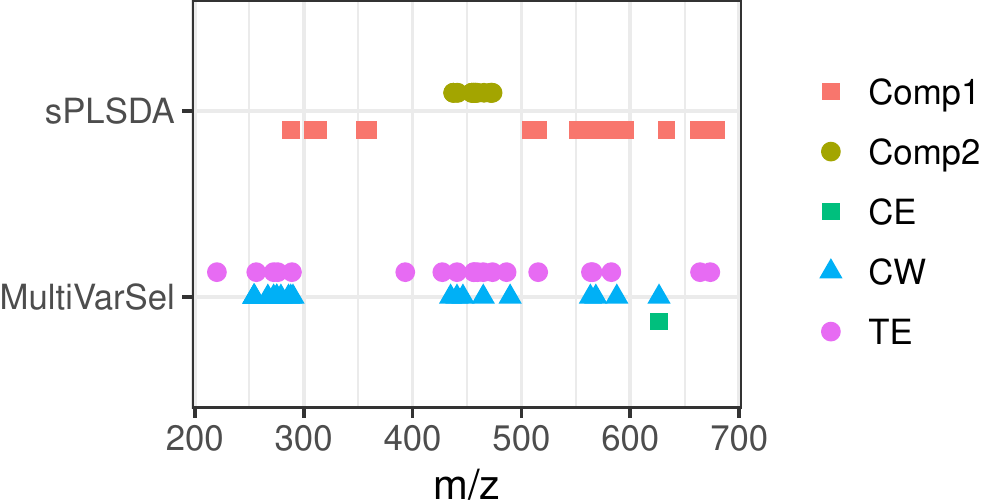}
\caption{Comparison of the metabolites selected by our approach \textsf{MultiVarSel} and by sPLS-DA.\label{fig:metabsel}}
\end{center}
\end{figure}

Figure \ref{fig:metabsel} displays the positions of the metabolites selected by our approach and sPLS-DA. We can see from this figure that out of the 39 selected 
metabolites, 6 metabolites are selected by both sPLS-DA and our methodology. The major difference between these two variable selection techniques is that
our method selects metabolites having a ratio $m/z$ smaller than 300 whereas the metabolites chosen by sPLS-DA lie within the range 300-400 $m/z$.

\begin{figure}[!h]
\begin{center}
\includegraphics[scale=0.8]{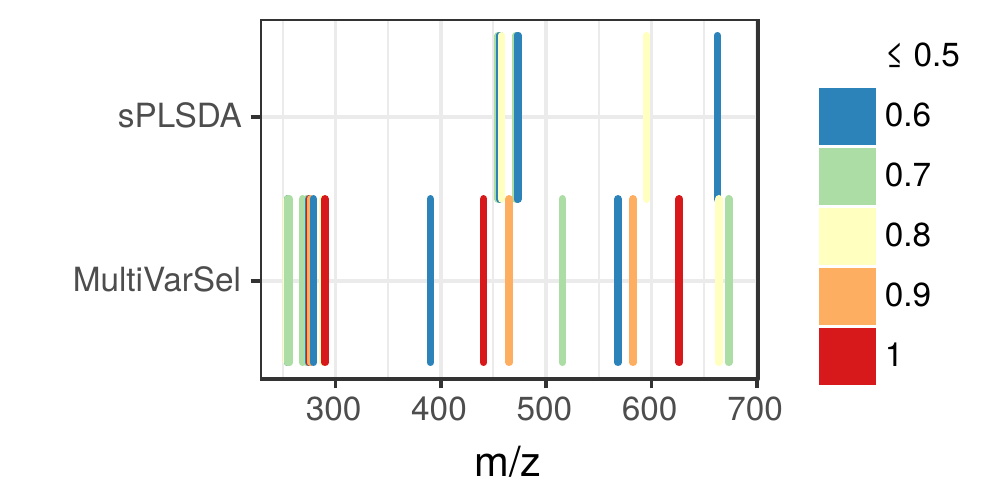}
\caption{Frequencies of the metabolites selected by sPLS-DA and our approach \textsf{MultiVarSel}.\label{fig:freq}}
\end{center}
\end{figure}

In order to further compare our methodology with sPLS-DA, we first propose to assess the stability of the selected variables (or metabolites). For this purpose, we 
performed 10 bootstrap resamplings of our original data and we compared the variables selected by both approaches. The results are displayed in
Figure \ref{fig:freq}  and in Table \ref{table:bootstrap_er_plsda}. Figure \ref{fig:freq} displays the frequencies at which each metabolite
has been selected by the two methods. We can see from this figure that the highest selection frequency of sPLS-DA is around 0.8 and that a lot of variables have a selection frequency smaller than 0.5. 
Moreover, we can see from Table \ref{table:bootstrap_er_plsda} which provides the number of metabolites which have been selected once (first row), 
twice (second row)..., that our approach selects 4 metabolites with a frequency equal to 1 which does not occur for sPLS-DA. Hence, from this point of view, our approach is
 more stable than sPLS-DA.

\begin{table}[h!]
    \begin{tabular}{ccc}
    \hline
Nb of selection & Nb of selected metabolites & Nb of selected metabolites \\
  & by sPLS-DA &  by \textsf{MultiVarSel} \\
\hline
    1 & 117 &  143\\
  2 &  41& 54\\ 
  3 &  26 & 24\\ 
  4 &   4 & 20\\ 
  5 &   8 & 15\\ 
  6 &   5 & 6\\ 
  7 &   3 & 6\\ 
  8 &   2 & 3\\ 
  9 & 0  & 3\\ 
10 & 0 & 4\\
    \hline
    \end{tabular}
\caption{Number of times the different metabolites have been selected by sPLS-DA and our approach \textsf{MultiVarSel}.\label{table:bootstrap_er_plsda}}
\end{table}

Finally, we compare hereafter the set of variables provided by sPLS-DA
and our approach from the classification error point of view. Since our method is not designed for yielding a classification we give the selected variables to PLS-DA in order to obtain 
such a classification.  The estimation of the classification error rates are then obtained by using a 10-fold cross-validation.
The corresponding results are displayed in Tables \ref{table:CV_er_splsda} and \ref{table:CV_er_our_plsda}. We observe that the classification error rates of our approach
are on a par with those of sPLS-DA.

\begin{table}[ht]
\centering
\begin{tabular}{rrrr}
  \hline
 & TE & CE & CW \\ 
  \hline
TE & 0.92 & 0.33 & 0.00 \\ 
  CE & 0.08 & 0.67 & 0.00 \\ 
  CW & 0.00 & 0.00 & 1.00 \\ 
   \hline
\end{tabular}
\caption{Classification error rates for sPLS-DA.\label{table:CV_er_splsda}}
\end{table}

\begin{table}[ht]
\centering
\begin{tabular}{rrrr}
  \hline
 & TE & CE & CW \\ 
  \hline
TE & 0.92 & 0.22 & 0.00 \\ 
  CE & 0.08 & 0.67 & 0.00 \\ 
  CW & 0.00 & 0.11 & 1.00 \\ 
   \hline
\end{tabular}
\caption{Classification error rates for our approach coupled with PLS-DA.\label{table:CV_er_our_plsda}}
\end{table}

We observe from these different investigations that our approach provides similar results as sPLS-DA in terms of classification even if our approach was not designed for this purpose and that 
it yields more stable variables (metabolites) than sPLS-DA for characterizing the different classes.


\section{Conclusion}

In  this paper,  we  proposed  a novel  approach  for analyzing  LC-MS
metabolomics data by introducing a new Lasso-type approach taking into
account the dependence that may exist  between the columns of the data
matrix.   Our approach  is  implemented in  the R  package \textsf{MultiVarSel}
which is available from the The Comprehensive R Archive Network (CRAN).   
In the course  of this study, we  have shown
that our method  has two main features. Firstly, it  is very efficient
from a statistical point of view  for selecting a restricted number of
stable metabolites   characterizing each level   of   the    factor   of
interest. Secondly,  its very low  computational burden makes  its use
possible on very large LC-MS metabolomics data.

\section*{Acknowledgements}

This    project    has    been    funded   by    La    mission    pour
l'interdisciplinarit\'e du  CNRS in the  frame of the  DEFI ENVIROMICS
(project AREA).  The  authors thank the Mus\'ee  Fran\c cois Tillequin
for providing the samples from the Guibourt Collection.


\subsection*{Appendix A}

Let $vec(\boldsymbol{A})$ denote the vectorization of the matrix $\boldsymbol{A}$ formed by stacking the columns of 
$\boldsymbol{A}$ into a single column vector. Let us apply the $vec$ operator to Model (\ref{eq:model:matriciel}), then
$$
vec(\boldsymbol{Y})=vec(\boldsymbol{X}\boldsymbol{B}+\boldsymbol{E})=vec(\boldsymbol{X}\boldsymbol{B})+vec(\boldsymbol{E}).
$$
Let $\mathcal{Y}=vec(\boldsymbol{Y})$,  $\mathcal{B}=vec(\boldsymbol{B})$ and $\mathcal{E}=vec(\boldsymbol{E})$.
Hence,
$$
\mathcal{Y}=vec(\boldsymbol{X}\boldsymbol{B})+\mathcal{E}=(\textrm{I}_q\otimes\boldsymbol{X})\mathcal{B}+\mathcal{E},
$$
where we used that \cite[Appendix A.2.5]{mardia:kent:1979}
$$
vec(AXB) = (B'\otimes A) vec(X).
$$
In this equation, $B'$ denotes the transpose of the matrix $B$.
Thus,
$$
\mathcal{Y}=\mathcal{X}\mathcal{B}+\mathcal{E},
$$
where $\mathcal{X}=\textrm{I}_q\otimes\boldsymbol{X}$ and $\mathcal{Y}$, $\mathcal{B}$ and $\mathcal{E}$ are vectors of size $nq$, $pq$ and $nq$, respectively.

\subsection*{Appendix B}

Let us apply the $vec$ operator to Model (\ref{eq:modele:blanchi_est})
where $\boldsymbol{\Sigma}_q^{-1/2}$ is
replaced by $\widehat{\boldsymbol{\Sigma}}_q^{-1/2}$, then
\begin{multline*}
vec(\boldsymbol{Y}\widehat{\boldsymbol{\Sigma}}_q^{-1/2})
=vec(\boldsymbol{X}\boldsymbol{B}\widehat{\boldsymbol{\Sigma}}_q^{-1/2})
+vec(\boldsymbol{E}\widehat{\boldsymbol{\Sigma}}_q^{-1/2})\\
=((\widehat{\boldsymbol{\Sigma}}_q^{-1/2})'\otimes \boldsymbol{X})vec(\boldsymbol{B})
+vec(\boldsymbol{E}\widehat{\boldsymbol{\Sigma}}_q^{-1/2}).
\end{multline*}
Hence,
$$
{\mathcal{Y}}={\mathcal{X}}\mathcal{B}+{\mathcal{E}},
$$
where ${\mathcal{Y}}=vec(\boldsymbol{Y}\widehat{\boldsymbol{\Sigma}}_q^{-1/2})$, 
${\mathcal{X}}=(\widehat{\boldsymbol{\Sigma}}_q^{-1/2})'\otimes \boldsymbol{X}$ 
and ${\mathcal{E}}=vec(\boldsymbol{E}\widehat{\boldsymbol{\Sigma}}_q^{-1/2})$.


\end{document}